\documentstyle[mnextra]{mn} 
\input{epsf}
\def\twodF{{2dFGRS}} 
\def\areatwodF{1094.727   +     740.3251  + 3.16557*100}
\def\areatwodF{2151.6}
\def\eg{{e.g.\ }}

\def \etal {et al.\ }
\def \ie {{\rm i.e.\ } }
\def \Mstar {M$^\star_{\rm b_J}$} 
\def \phistar{$\Phi^\star$}
\def \btheta {{\bmath \theta}}
\def \bj {{\rm b_J}}

\def \a {{\rm a}}
\def \b {{\rm b}}
\def \c {{\rm c}}
\def \d {{\rm d}}

\def \g {{\rm g}}
\def \r {{\rm r}}

\def \deg {$^{\rm o}$}
\def \kms {{\rm km\, s^{-1}}} 
\def \logh {{\log_{10} h}} 
\def \date {May 2001}
\def \nztot {153\,986}     
\def \phistarval{1.64}
\def \phistarvaldecon{1.61}
\def \Mag{M_\bj - 5 \logh} 
\def \K{{\rm K}_{\rm S}}
\def \J{{\rm J}}
\def \R{{\rm R}}
\def \B{{\rm B}}
\def \V{{\rm V}}

\nobrackets     

\begin{document}

\title[2dFGRS $\bj$-band Luminosity Function]{The 2dF Galaxy Redshift Survey: 
The b$_{\bf J}$-band galaxy luminosity function and survey selection function
\vspace*{-0.3 truecm} }

\author[P. Norberg et al.] {
\parbox[h]{\textwidth}{Peder Norberg$^1$, Shaun Cole$^1$, 
Carlton M. Baugh$^1$, Carlos S. Frenk$^1$, Ivan Baldry$^{10}$,
Joss Bland-Hawthorn$^2$, 
Terry Bridges$^2$,
Russell Cannon$^2$, Matthew Colless$^3$, Chris Collins$^4$, 
Warrick Couch$^5$, Nicholas J.G. Cross$^6$,
Gavin Dalton$^7$, Roberto De Propris$^5$,
Simon P. Driver$^{6,3}$, 
George Efstathiou$^8$, Richard S. Ellis$^9$, Karl Glazebrook$^{10}$, 
Carole Jackson$^3$,
Ofer Lahav$^8$, Ian Lewis$^7$, Stuart Lumsden$^{11}$, 
Steve Maddox$^{12}$, Darren Madgwick$^8$,
John A. Peacock$^{13}$, Bruce A. Peterson$^3$,
Will Sutherland$^{13}$, Keith Taylor$^2$ (The 2dFGRS Team)}
\vspace*{6pt} \\ 
$^1$Department of Physics, University of Durham, Science Laboratories, South 
Road, Durham DH1 3LE, United Kingdom \\
$^2$Anglo-Australian Observatory, P.O. Box 296, Epping, NSW 2121, Australia \\
$^3$Research School of Astronomy \& Astrophysics, The Australian National 
University, Weston Creek, ACT 2611, Australia \\
$^4$Astrophysics Research Institute, Liverpool John Moores University, Twelve 
Quays House, Egerton Wharf, Birkenhead, L14 1LD, UK \\
$^5$Department of Astrophysics, University of New South Wales, Sydney, 
NSW2052, Australia \\
$^6$School of Physics and Astronomy, North Haugh, St Andrews, Fife, KY16 9SS,
United Kingdom \\
$^7$Department of Physics, Keble Road, Oxford OX1 3RH, United Kingdom \\
$^8$Institute of Astronomy, University of Cambridge, Madingley Road, 
Cambridge CB3 0HA, United Kingdom \\
$^9$Department of Astronomy, California Institute of Technology, Pasadena, 
CA 91125, USA \\
$^{10}$Department of Physics \& Astronomy, Johns Hopkins University, 3400 
North Charles Street Baltimore, MD 21218--2686, USA \\
$^{11}$Department of Physics \& Astronomy, E C Stoner Building, Leeds LS2 9JT,
United Kingdom \\
$^{12}$School of Physics and Astronomy, University of Nottingham, University 
Park, Nottingham, NG7 2RD, United Kingdom \\
$^{13}$Institute of Astronomy, University of Edinburgh, Royal Observatory, 
Edinburgh EH9 3HJ, United Kingdom \\
\vspace*{-1.0 truecm}
}

\maketitle

\begin{abstract}
We use more than 110\,500 galaxies from the 2dF galaxy redshift survey
(2dFGRS) to estimate the $\bj$-band galaxy luminosity function at
redshift $z=0$, taking account of evolution, the distribution of
magnitude measurement errors and small corrections for incompleteness
in the galaxy catalogue. Throughout the interval
$-16.5>M- 5 \logh > -22$, the luminosity function is accurately
described by a Schechter function with $M^\star_\bj -5 \logh =-19.66\pm0.07$, $\alpha=-1.21\pm0.03$ and
\phistar$=(1.61\pm0.08)\times 10^{-2} h^3$Mpc$^{-3}$, giving
an integrated luminosity density of 
$\rho_L=(1.82\pm0.17)\times 10^8 h$L$_\odot$~Mpc$^{-3}$
(assuming an $\Omega_0=0.3$, $\Lambda_0=0.7$ cosmology).  
The quoted errors have
contributions from the accuracy of the photometric zeropoint, large
scale structure in the galaxy distribution and, importantly, from the
uncertainty in the appropriate evolutionary corrections. Our
luminosity function is in excellent agreement with, but has much
smaller statistical errors than an estimate from the Sloan Digital Sky
Survey (SDSS) data when the SDSS data are accurately translated to the
$\bj$-band and the luminosity functions are normalized in the same
way.  We use the luminosity function, along with maps describing the
redshift completeness of the current 2dFGRS catalogue, and its weak
dependence on apparent magnitude, to define a complete 
description of the 2dFGRS selection function. Details and tests
of the calibration of the 2dFGRS photometric parent catalogue
are also presented.
\end{abstract}
\begin{keywords}
galaxies: luminosity function -- selection function -- 2dF galaxy redshift survey (2dFGRS) -- mock catalogues 
\vspace*{-0.5 truecm}
\end{keywords}

\section{Introduction}

The galaxy luminosity function (LF), which gives the abundance of galaxies 
as a function of their luminosity, is one of the most fundamental properties 
of the galaxy distribution. The accuracy with which it is known
has improved steadily as the size of the redshift surveys used  
to determine it has grown 
(\eg \cite{swml,loveday,marzke,lin,zucca,ratcliffe,folkes,blanton,madgwick01}).
Here, we present an estimate of the $\bj$-band luminosity function from 
the 2dF Galaxy Redshift Survey (2dFGRS) which is currently the largest 
galaxy redshift survey in existence. The luminosity function is an important
statistic in its own right and understanding how it arises is 
a major goal of models of galaxy formation
(\eg White \& Frenk \shortcite{wf}; Katz, Hernquist \& Weinberg \shortcite{katz92};
Kauffmann, White \& Guiderdoni \shortcite{kauffmann};
Cole \etal \shortcite{cole94}, \shortcite{cole2k};
Somerville \& Primack \shortcite{somerville}; Pearce \etal
\shortcite{pearce01}).
Also, to exploit the 2dFGRS fully, it is important to have an accurate model
of the luminosity function so that the selection function of the
survey can be determined. This is a vital ingredient in analysing
all aspects of galaxy clustering using the survey.

 This paper presents an estimate of the overall $\bj$-band galaxy
luminosity function. This estimate takes account of k-corrections
(which result from the redshifting of the wavelength range covered
by the $\bj$ filter)
and also average evolutionary corrections.  
We also include the effects of photometric errors and small
corrections for incompleteness in the survey. We demonstrate
that the dependence of the incompleteness on surface brightness is small,
but we do not include any explicit surface brightness corrections. These 
will be discussed in Cross \etal (\shortcite{cross01}).
The analysis presented here is complementary to that
in Madgwick \etal (\shortcite{madgwick01}) and the earlier analysis in Folkes
\etal (\shortcite{folkes}). In these cases a subset of the
2dFGRS data were analyzed with the primary aim of establishing how the
luminosity function depends on spectral type. These papers did not
apply evolutionary corrections since they were not attempting to model
the full selection function of the survey. We compare and discuss our
result in relation to these and other recent determinations of the
luminosity function, including that from the Sloan Digital Sky Survey
(SDSS).  We also compare estimates for different regions of the survey
to test the uniformity of the catalogue and our model assumptions.
Throughout, we use mock galaxy catalogues constructed from the Hubble
Volume N-body simulations (\cite{evrard99,evrard01}) 
in order to check our
methods and to assess the influence of large scale structure upon our
results.  We also use the estimated luminosity function and our
modelling of the survey selection limits and completeness to produce a
complete description 
of the 2dFGRS selection function in
angle, redshift and apparent magnitude.  
The predictions of this selection function
are compared with various properties of the real catalogue including
the galaxy number counts and redshift distributions.

The paper is divided into \ref{sec:concl} sections.  In
Section~\ref{sec:sample} we describe the relevant details of the
2dFGRS. The technical details of the calibration and accuracy of the APM
and 2dFGRS photometry are discussed in Section~\ref{sec:apm_cal}.
In Section~\ref{sec:sdss} we use independent data from the
SDSS to review the reliability of the 2dFGRS redshifts 
and investigate various aspects of the
completeness of the 2dFGRS photometric and redshift catalogues.  
This is again a technical section and the general reader may 
wish to skip these sections.
In Section~\ref{sec:k+e} we describe how we model the galaxy k+e corrections.
In Section~\ref{sec:mocks} we briefly
describe a set of mock catalogues, which we use both to test our
implementation of the luminosity function estimators and to assess the
effects of large-scale structure.  We present a series of luminosity function
estimates in Section~\ref{sec:lf}, where we compare results for
different regions and subsets of the survey.  In Section~\ref{sec:lf.norm}
we examine the 2dFGRS number counts that we use to normalize our
LF estimates and compare them to counts from the
SDSS. Our normalized estimate of the 2dFGRS luminosity function is
presented in Section~\ref{sec:lf2df}.
We compare our results with independent LF estimates in
Section~\ref{sec:lf.compare}.  In Section~\ref{sec:selfun} we use our
best estimate of the 2dFGRS LF, together with the
description of the survey magnitude limits and completeness, to
construct a model of the survey selection function. From this we
extract the expected redshift distribution which we compare with 
those of the real survey and mock catalogues.
We discuss our results and present our conclusions in Section~\ref{sec:concl}.

\section{The 2\lowercase{d}F Galaxy Redshift Survey}
\label{sec:sample} 

\begin{figure*}
\epsfxsize=18.5 truecm \epsfbox{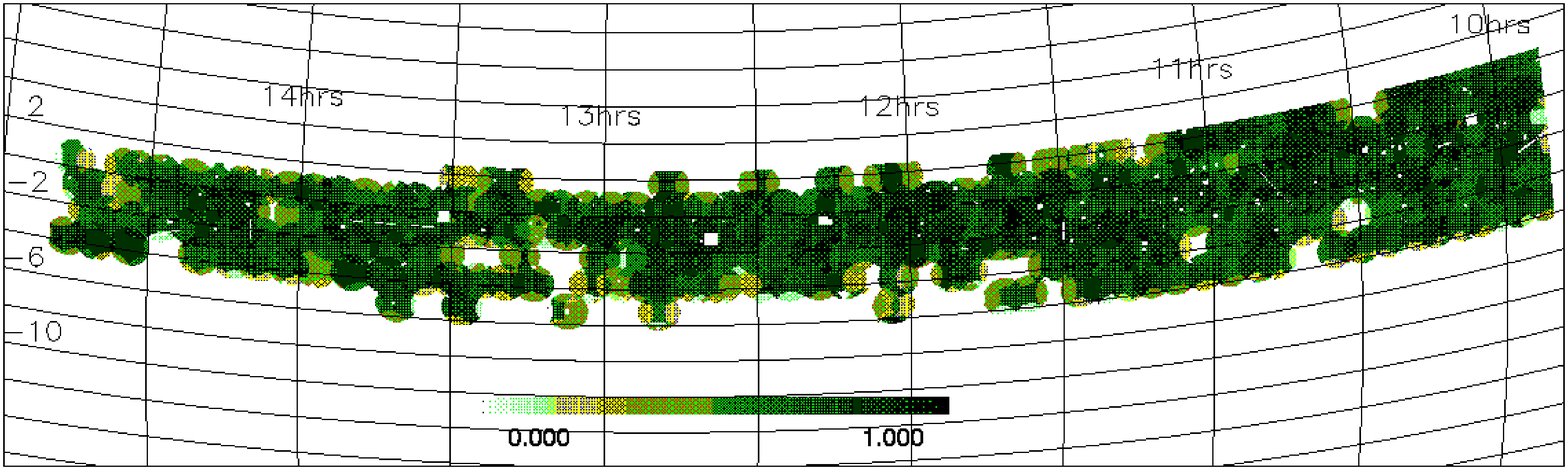}
\epsfxsize=18.5 truecm \epsfbox{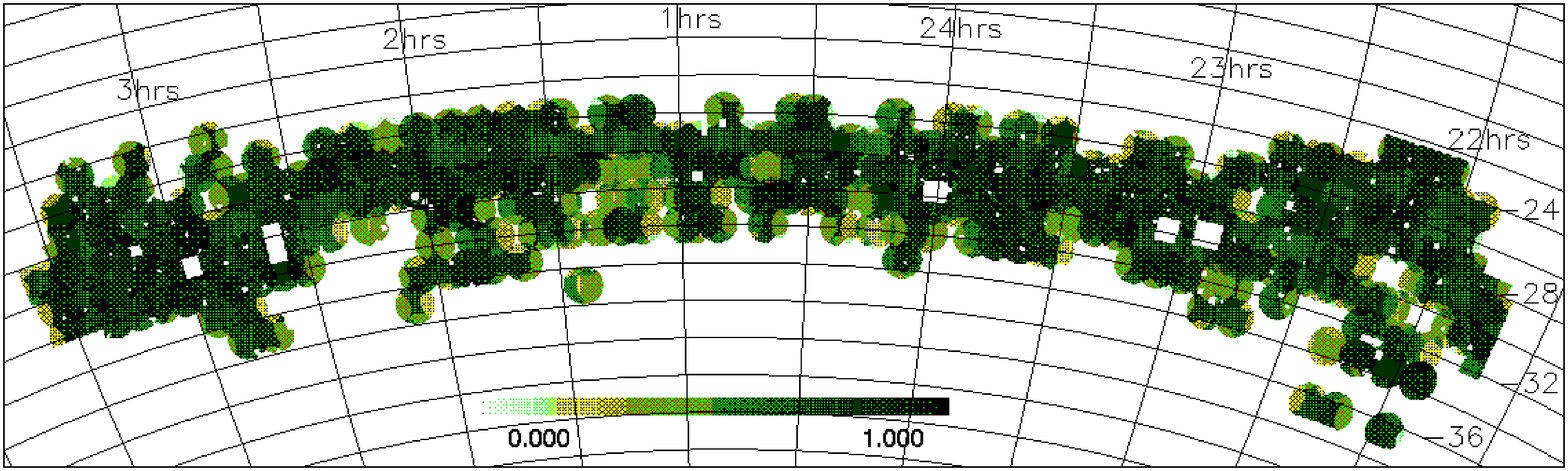}
\caption{The sky coverage of the 2dFGRS dataset analysed in this paper.
This dataset includes galaxy redshifts from all fields observed before \date\ 
that have a redshift completeness greater than 70\%. 
As the fields overlap and many are still to be observed,
the completeness varies across the sky.
The quantity represented by the grey-scale is the sector 
redshift completeness, $R(\btheta)$,
defined in Appendix~\ref{app:maps}.
}
\label{fig:mask}
\end{figure*}

The \twodF\ is selected in the photographic $\bj$ band from the APM
galaxy survey (Maddox \etal \shortcite{apmI}, \shortcite{apmII},
\shortcite{apmIII}) and subsequent extensions to
cit that include a region in the northern galactic cap
(\cite{2dfinput}).  The survey covers approximately \areatwodF~deg$^2$
in two broad declination strips. The larger of these is centred on
the South Galactic Pole (SGP) and approximately covers
$-22^\circ\negthinspace.5$ $>$ $\delta$ $>$
$-37^\circ\negthinspace.5$, $21^{\rm h}40^{\rm m}$ $<$ $\alpha$ $<$
$3^{\rm h}40^{\rm m}$; the smaller strip is in the northern galactic
cap and covers $2^\circ\negthinspace.5$ $>$ $\delta$ $>$
$-7^\circ\negthinspace.5$, $9^{\rm h}50^{\rm m}$ $<$ $\alpha$ $<$
$14^{\rm h}50^{\rm m}$.  In addition, there are a number of pseudo-randomly
located circular 2-degree fields scattered across the full extent of
the low extinction regions of the southern APM galaxy survey.  There
are some gaps in the \twodF\ sky coverage within these boundaries due
to small regions that have been excluded around bright stars and
satellite trails. The aim of the \twodF\ is to measure the redshifts of all the
galaxies within these boundaries with extinction-corrected $\bj$
magnitudes brighter than 19.45.  As described in Colless \etal
(\shortcite{colless01}), this is attempted by dividing the target
galaxies among a series of overlapping 2\deg\ diameter fields. The
degree of overlap of the fields is such that the number of targets
assigned to each field is no greater than the 400~fibres that the 2dF
instrument uses to obtain spectra for each target simultaneously. When
all these 2\deg\ fields have been observed, in early 2002, close to
250\,000 galaxy redshifts will have been measured.

In this paper we use the \nztot\ redshifts obtained prior to \date\ 
in the main NGP and SGP strips.
This sample covers a large fraction of the full 2dFGRS area, but
as shown in Fig.~\ref{fig:mask},
within this area the sampling rate varies with position on the sky.
This is a direct consequence of some of the overlapping 2\deg\ fields
having not yet been observed and so is well understood and can be
accurately modelled (see section~8 of Colless \etal \shortcite{colless01}).

For accurate statistical analysis of the 2dFGRS it is essential to
understand fully the criteria that define its parent photometric
galaxy catalogue and also the spatial and magnitude dependent
completeness of the redshift survey. In the remainder of this section
we describe the calibration and photometric accuracy of the parent
galaxy catalogue. The accuracy and completeness of the redshift survey
have been quantified in the survey paper Colless \etal
(\shortcite{colless01}).  Later, in Section~\ref{sec:sdss}, we complement
the analysis of Colless \etal (\shortcite{colless01}) and our
description of the 2dFGRS photometry by making a direct comparison
with data from the overlapping Early Data Release (EDR) of the Sloan
Digital Sky Survey (SDSS).

   A detailed description of the calibration of the original APM
catalogue can be found in Maddox \etal (\shortcite{apmII} hereafter
APMII).  Here we start, in \S\ref{sec:apm_cal}, by reviewing the important
steps in this process, referring the reader to sections of APMII
where many more details can be found along with various tests of the
assumptions made in modelling the photometry.  In \S\ref{2df_cal}, we
describe the additional calibrating data and dust corrections that
were used to define the input catalogue used to select objects for the
2dFGRS (see also \cite{2dfinput}). 
Finally, in \S\ref{100k_cal}, we describe how recently
available CCD data and revised dust maps have been used to define the
2dFGRS magnitudes which were made public in the 100k release and
which we adopt throughout this paper.

\subsection{APM Photometric Calibration}
\label{sec:apm_cal}

The APM measures photographic density and for each image
calculates the integrated density and area within an isodensity
contour.  If the images are not saturated, these are equivalent to an
uncalibrated isophotal magnitude and corresponding isophotal
area. These are converted to uncalibrated raw total magnitudes by
modelling the intensity profile of each image as a gaussian. It is
argued that this is a sufficiently accurate assumption as the observed
profiles of the fainter galaxies are dominated by gaussian seeing,
while the isophotal correction is small for the bright galaxies (APMII
\S2.1).

    The isodensity threshold of the APM images corresponds roughly
to a surface brightness of $25.0\,$mag arcsec$^{-2}$ 
(Shao \etal \shortcite{apmsb}), but varies 
significantly within each UKST field. The main causes of this
variation are geometric vignetting (variation of the effective
area of the telescope with off axis angle) and desensitization of
the hyper-sensitized photographic emulsion, which varies both
systematically and randomly with field position (APMII~\S2.2).
If the intrinsic sky brightness is uniform across the field
of view then variations in the measured 
sky brightness can be used to estimate the 
variation of the sensitivity across the plate. In APMII~\S2.3,
a model is developed to correct the raw total magnitudes
using such maps of the measured sky brightness across each plate.
This model assumes that the magnitudes being corrected are total
magnitudes, that the true sky brightness is uniform across the field
of view and that the UKST plates are of uniformly high quality.
These corrections do not take account of saturation 
effects which are expected to become significant for galaxies brighter
than $\bj=16$ and may affect fainter high surface brightness ellipticals.      

After applying the corrections described above the resulting
field-corrected total magnitudes, $m_{\rm FC}$, are consistently
defined over an individual UKST plate. However, the zeropoint
and possibly the  non-linearity can vary from plate to plate.
Matched images on the substantial overlaps between the plates are used
to find a set of transformations that bring the magnitudes on all
the plates onto a common scale (APMII~\S3). Linear transformations 
of the form
\begin{equation}
      m_{\rm A}   = a_i + b_i m_{\rm FC}
\label{APMFC}
\end{equation}
were adopted to express the matched APM magnitudes, $m_{\rm A}$
in terms of the field-corrected magnitudes. 
Here, for plate $i$,  $a_i$ is the zeropoint offset 
and $b_i$ a term that can correct for residual non-linearity.
The degree of non-linearity in the original APM plates is quite small
with an rms variation $\langle(b_i-1)^2\rangle^{1/2}=0.05$.

This procedure does not constrain the overall zeropoint or linearity
of the survey. This global calibration was done using a limited number
of $\B$ and $\V$-band CCD sequences that were spread reasonably
uniformly across the survey area (APMII~\S4).  Total CCD magnitudes
were measured and converted into the $\bj$-band using $\bj=\B
-0.28(\B-\V)$ (Blair \& Gilmore \shortcite{bj}).
Then a calibration curve of the form
\begin{equation}
 \bj = a_{\rm global} + b_{\rm global}\, m_{\rm A} + c_{\rm global}\, m_{\rm A}^2
\label{APMGLOBAL}
\end{equation}
was determined by minimising the residuals between the final
APM $\bj$ and CCD total magnitude. A plot of the resulting calibration
curve (both with and without the quadratic term) is shown in  figure~1 of
Maddox \etal (\shortcite{apm_counts}).
The CCD sequences were also used as a check of the zeropoints
determined by plate matching (APMII~\S4.2).

It is important to realize that while the APM magnitudes are based on
measurements which are approximately equivalent to an isophotal
magnitude (with a somewhat uncertain and varying isophote), the APM
correction and calibration procedure converts these into estimates of
{\it total} magnitudes. The result should be that for galaxies in each
interval of apparent magnitude the mean calibrated APM magnitude
equals the mean total magnitude of the calibrating CCD data.
The scatter about this mean relation is approximately $0.16$ magnitudes,
this being driven mainly by the inaccuracy of photographic photometry,
but also it is to be expected that within this scatter the residuals 
will correlate with surface brightness and possibly other properties
of the galaxy images.

\subsection{Calibration of 2dFGRS input catalogue}
\label{2df_cal}

In 1994, when the input catalogue of the 2dFGRS was specified,
substantially more CCD data was available 
than when the original APM catalogue was calibrated.
Also new, improved plates and/or APM scans were available
for a few UKST fields. Consequently  both the plate matching and the
final global calibration steps described above were redone using 
the new data. 
This time the parameters $b_i$ in equation (\ref{APMFC}) were
kept fixed at unity. The reason for this was that in the original
APM survey the deviations from linearity were quite
small; also, for the purposes of selecting the $\bj<19.45$ catalogue
it was not necessary to have accurate bright magnitudes. The galaxies
used in the plate matching were selected to have magnitudes in the
range $19.5<\bj<20.5$, close to the 2dFGRS magnitude limit.

The new CCD data was provided by Jon Loveday \& Simon Lilly (private
communication) and consists of 330 $10'\times 10'$ fields in $\B$ and
$\R$ centred on galaxies selected from the Stromlo-APM catalogue
(Loveday \etal \shortcite{sapm}).  The data was taken using the CTIO
1.5m and total magnitudes were determined using using Sextractor
(\cite{bertin}). The integration times were 2~minutes in $\R$ and
5~minutes in $\B$, leading to completeness limits of $\B\sim 21$ and
$\R\sim 20$.  This yields $13\,162$ images brighter than $\bj=21$. Of
these, $1\,760$ matched with galaxies in the APM data and these were
used to determine the global calibration of APM magnitudes (\ie
determine the parameters of equation~\ref{APMGLOBAL}).  This new
calibration was slightly different to the original APM calibration,
both in terms of linearity correction and zero-point.

To accommodate an efficient observing strategy for the 2dFGRS it was
necessary to extend the APM survey. Two overlapping strips of UKST
plates in the region $9^{\rm h}<\alpha<15^{\rm h}$, 
$-7^\circ\negthinspace.5<\delta<3^\circ\negthinspace.5$, 
close to the NGP, were reduced using the standard APM galaxy 
survey procedures.  For this area significant quantities of
calibrating CCD data 
were not directly available. However, APM scans of the same UKST plates 
(reduced using a somewhat different suite of software) had been
calibrated using CCD data by Raychaudhury \etal (\shortcite{somak}).  
We had access to these calibrating catalogues, but not the original
CCD data. We, therefore, set our global
calibration by minimising the galaxy by galaxy residuals between our
magnitudes and the calibrated magnitudes given by Raychaudhury \etal
(\shortcite{somak}).

The final step in the preparation of the magnitudes used to select the
2dFGRS parent catalogue was to apply extinction corrections. 
For this we used a new, high resolution dust map supplied to us
by David Schlegel, in advance of the slightly modified
map that was subsequently published in Schlegel, Finkbeiner \& Davis 
(\shortcite{dust}).

\begin{figure*}
\epsfxsize=15.5 truecm \epsfbox[0 510 520 748]{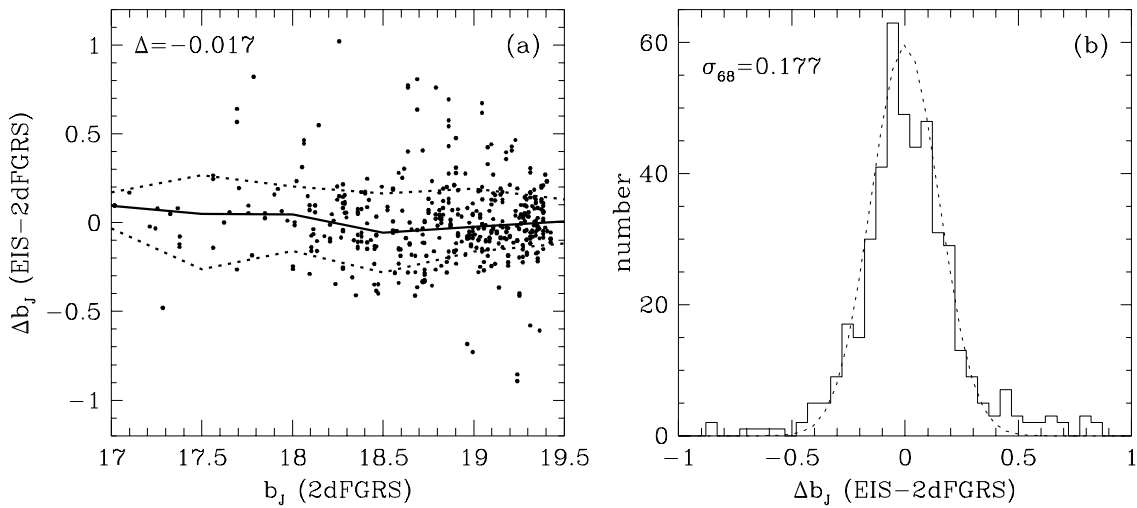}
\caption{Comparison of 2dFGRS photographic $\bj$ magnitudes
with CCD magnitudes from EIS Patch B.
Panel (a) is a scatter plot of the magnitude 
difference versus 2dFGRS magnitude and the solid and dotted lines show the 
magnitude dependence of the median, 16\% and 84\% quantiles of the 
distribution. The median magnitude difference, $\Delta$, 
for all the galaxies in the range $17<\bj<19.5$ is indicated on the panel. 
The distribution of magnitude differences with respect to this median is
shown as a histogram in panel (b).
The dotted curve, which describes the core of this 
distribution quite well, is a gaussian with $\sigma=0.15$ magnitudes. 
A robust estimate of the width of this distribution, $\sigma_{68}$, defined
such that $2\sigma_{68}$ spans 68\% of the distribution, is also indicated 
on the panel. }
\label{fig:eis_mags}
\end{figure*}

\subsection{100k 2dFGRS recalibration}
\label{100k_cal}

Here we describe the improvements that were made to the 2dFGRS
calibration and extinction corrections after target selection. These
improved magnitudes (as well as the original magnitudes on which
target selection was based) are available in the public 100k
Release\footnote{The data in 100k Release are publically available at
http://www.mso.anu.edu.au/2dFGRS/ } 
for the whole of the 2dFGRS parent catalogue.
In outline, two main changes were made:
\begin{enumerate}

\item Corrections for plate-to-plate nonlinearities. The original calibrating
CCD photometry concentrated on galaxies near $\bj\simeq 20$, so the linearity of
the photometry at bright magnitudes was hard to check. More recent CCD
data showed that significant nonlinearities were sometimes present.
The main tool used to correct these was the mean optical to infrared
colour on each plate between 2dFGRS and the 2MASS survey.

\item Improved estimates of the foreground extinction.
The original 2dFGRS selection, performed in 1994,
was based on magnitudes corrected
using a preliminary version of the Schlegel \etal (\shortcite{dust}) dust map.
The final version and calibration of this map differed slightly from 
the one originally adopted. On average,
the updated extinction corrections are larger than those initially adopted, by
a factor of approximately 1.3. This yields rms shifts in zero point of
0.02~magnitudes in the NGP, and half this in the SGP.

\end{enumerate}

In principle, variations in plate-to-plate nonlinearity might not
be expected to be a problem. These effects can be diagnosed when
the magnitudes in the overlap region between two different plates
are compared. The ideal case would then be a set of plates with
a consistent set of magnitudes in the overlaps, whose absolute zero point and
degree of linearity can be assessed by combining all the calibrating
CCD photometry. The practical difficulty is that the number of
bright galaxies for comparison in the overlaps can be small where
one of the plates is of lower quality, preventing an accurate
determination of the linearity of that plate. This turns out to
have been a significant problem in the NGP zone, where we used
fewer plates than in the SGP, and where the quality was less
homogeneous. Nevertheless, there were some nonlinearity problems
in the SGP also.

The direct measurement of plate nonlinearities requires large
CCD datasets. In the SGP, we were able to make comparison with the
results of the public ESO Imaging Survey (EIS) in the Patch B region 
(Prandoni \etal \shortcite{eisb})
and the Chandra Deep Field (Arnouts \etal \shortcite{eis}).
We also had access to data from the ESO--Sculptor Survey
(Arnouts \etal \shortcite{eso}).
The Patch B dataset is the largest, and showed the 2dFGRS
photometry to be consistent with linear in this plate (UKST 411);
there was significant nonlinearity in the Chandra Deep Field
(UKST 418), however.
The EIS datasets have accurately determined colour
equations, so we are able to make
an accurate determination of the colour equation
between $\bj$ and Johnson B \& V:
\begin{equation}
      \bj = \B - (0.267\pm0.019)(\B-\V),
\end{equation}
verifying our standard colour term of 0.28.

In the north, the main comparison was with the MGC survey (\cite{mgc}). 
This covers a $36\,$deg$^2$ strip between $10^{\rm h}<\alpha<15^{\rm h}$
and overlaps with 15 UKST plates in the 2dFGRS NGP strip
(a total of 5205 galaxies in common).
Data are available only in a blue filter ($\B_{\rm MGC}$) that
differs somewhat from the 2dFGRS $\bj$. The MGC thus cannot
yield an absolute calibration for 2dFGRS, but the accuracy of the
data (a limiting magnitude of $\B_{\rm MGC}\simeq 24$) means that
precise measures of nonlinearity and relative zero-points were possible.
For each UKST plate, we fitted a linear relation of the form
\begin{equation}
      \B_{\rm MGC}  =  a_i + b_i \bj,
\end{equation}
where $\bj$ is the observed 2dFGRS magnitude, prior to dust correction.
In this way we determined the non-linearity $b_i$ and
relative zeropoint $a_i$ of each of the 15 UKST plates.
Nonlinearities of up to $|\b_i -1|=0.2$ were measured (median 0.08).
That variations in linearity exist in the NGP strip is not
surprising.  As discussed in APMII~\S2.3, the model that was
constructed to correct the raw APM magnitudes for vignetting and
variations in plate sensitivity assumes that the UKST plates are of
uniform quality. In particular, it is assumed that the sky brightness
varies by only a small amount both across the field of view of each
plate and from one plate to another.
This assumption is less valid for the NGP strip, as the
UKST plates used here were less homogeneous than in the original APM survey.

Based on these results, nonlinearities must also exist for
plates where large CCD datasets are lacking. In the 
absence of this information, we resorted to the only uniform
all-sky source of digital galaxy photometry: the
near-infrared 2-Micron All-Sky Survey (\cite{2mass}). In particular,
we concentrated on the $\J$-band 2MASS magnitudes for 2MASS-detected
galaxies that are also in the 2dFGRS.
Although the 2MASS data are at a much longer
wavelength than the APM photometry, 
nevertheless comparing 2MASS and APM photometry proved to be very
useful.  The plate matching procedure used to define the 2dFGRS
magnitudes should be accurate at 
approximately $\bj=20$ (the typical magnitude of most of the
galaxies used in the matching), and this is verified by our
original calibrating CCD photometry. However, residual non-linearity
can result in significant plate-to-plate offsets at bright magnitudes.
As the median magnitude of the matched 2MASS--2dFGRS galaxies is
$\bj \simeq 17.5$, such offsets manifest themselves as variations in the
median $\bj-\J$ of the matched galaxies.
For this comparison, we used magnitudes corrected for extinction
using the most recent dust maps. 
Any offset ($\Delta_{\rm 2MASS}$) in this median colour over a 
single UKST plate with respect to the global average should thus
indicate an error in the bright 2dFGRS magnitudes, 
and hence a nonlinearity.
We were able to verify that this assumption worked well by comparison
with the 18 plates with direct nonlinearity measurements, and we
therefore assumed that measurements of $\Delta_{\rm 2MASS}$ could
be used to diagnose nonlinearities in all plates.
In practice,
rather than assuming that plate overlaps assured an exact
matching at some specific magnitude between 20 and 21, an empirical
approach was taken. The MGC data showed that both nonlinearity
and the relative zero-point offsets at 19.45 correlated with
$\Delta_{\rm 2MASS}$, so a single value for $\Delta_{\rm 2MASS}$ was
used to give an estimate for each of these quantities.

The practical problem with this strategy is that not all
plates have an accurate measurement of $\Delta_{\rm 2MASS}$.
The dispersion in $\bj-\J$ is approximately 0.5~magnitudes, so
several hundred 2MASS galaxies per plate are required to give a
sufficiently accurate measurement of the offset.
Some plates fail to satisfy this criterion, since the full 2MASS
survey is not yet available. We therefore proceeded as follows.
Examining the trend of $\Delta_{\rm 2MASS}$ 
with sky position showed strong evidence for an
approximately linear variation of $\Delta_{\rm 2MASS}$
with RA and declination in both NGP and SGP. 
It is quite understandable that the
plate-matching could allow a slow drift of the magnitude linearity in
this way; we therefore used a linear fit as the initial estimate of
$\Delta_{\rm 2MASS}$, and hence nonlinearity, in any given plate.
Where the measurement of $\Delta_{\rm 2MASS}$ was
significantly different from the position-dependent fit,
a Bayesian approach was adopted, as follows.
Consider the deviation of the exact value of
$\Delta_{\rm 2MASS}$ with respect to the fit
-- call this $d_{\rm t}$, and let it have 
a prior probability distribution $P(d_{\rm t})$. 
We have an estimate of $d_{\rm t}$ from 
the limited 2MASS data that actually exist -- call this $d_{2\rm m}$. 
We are interested in the probability of $d_t$ given $d_{2\rm m}$, 
for which Bayes' theorem says 
 \begin{equation}
       P(d_{\rm t} | d_{2\rm m}) \propto P(d_{\rm t}) P(d_{2\rm m} |
       d_{\rm t}) .
 \end{equation}
This probability can be maximized to give a preferred value 
of $d_{\rm t}$ for a given $d_{2\rm m}$, 
taking into account the known statistical errors
on $d_{2\rm m}$. An estimate of the prior $P(d_{\rm t})$ 
was obtained from the
plates with very accurate values of $d_{2\rm m}$ and/or those with CCD
data.
This procedure allows us to interpolate smoothly between accepting
$\Delta_{\rm 2MASS}$ where it is well defined, and using the
position-dependent fit where the 2MASS data are sparse on an individual
plate. This yields a best estimate of $\Delta_{\rm 2MASS}$ for each
plate, and hence an estimate of the linearity of the initial 2dFGRS
photometry on that plate.
Of course, direct measurements of nonlinearity from CCD
data are to be preferred where they exist, and the direct
results from the MGC and EIS comparisons were used to recalibrate the
relevant plates, without reference to the 2MASS results.

Thus for the magnitudes given in the 100k release the use of
2MASS and MGC data has augmented the matching of plates that was done using
the overlaps. In the SGP the corrections to the magnitudes are
modest. The rms variation in linearity is
$\langle(b-1)^2\rangle^{1/2}=0.034$ and the rms shift of the zeropoint
at $\bj=19.45$ is $0.043$~magnitudes. In the NGP, which was
constructed from a less homogeneous set of UKST plates, the changes
are much more significant:
the rms variation in linearity is
$\langle(b-1)^2\rangle^{1/2}=0.106$ and the induced rms shift of the 
zeropoint at $\bj=19.45$ is $0.123$~magnitudes.

After applying these nonlinear transformations, the corrected magnitudes
should now be on a consistent scale on all plates in both SGP and NGP. The
overall zeropoint will still be that set at the faint end
by our initial CCD calibration. As an external absolute check
of the zeropoint, we compared with the optical CCD data from the EIS
Patch B (Prandoni \etal \shortcite{eisb}),
the EIS Chandra Deep Field (Arnouts \etal \shortcite{eis}), 
and the ESO-Sculptor survey (Arnouts \etal \shortcite{eso}), since these
are the largest datasets, and have the
best characterized colour equations. The
mean offset of these three fields with respect to the
original APM magnitudes was $0.03$~magnitudes, in the sense that the
original magnitudes were too faint.
This shift was applied, placing our magnitudes
on average on the same zeropoint as the EIS. The standard deviation
of the three offsets was $0.035$~magnitudes, which is in fact smaller
than the $0.07$ plate-to-plate dispersion that we would expect
over a large sample.
Therefore, it is conservative to assume that the rms uncertainty
in the 2dFGRS overall zeropoint is $0.07/\sqrt{3} =0.04$ magnitudes,
assuming no error in EIS. The internal tests of
Prandoni \etal (\shortcite{eisb}) suggest that any systematic
errors in the EIS calibration are smaller than this figure.
The EIS and calibrated 2dFGRS magnitudes are compared in 
Fig.~\ref{fig:eis_mags}. The rms error in an 
individual galaxy magnitude is approximately 0.15.
The effect of these recalibrations, and the change
in the dust correction, is
shown in figures~13 and~14 of Colless \etal (\shortcite{colless01}).

\section{Comparison with the SDSS}
\label{sec:sdss}

Here we complement the description 
given above and in the survey construction papers (\cite{colless01,2dfinput})
by making a direct comparison of the 2dFGRS catalogue with the 
Early Data Release (EDR) of the Sloan Digital Sky Survey (SDSS).
The two datasets have approximately 30\,000 galaxies in common of which
about 10\,000 have redshift measurements in both surveys.
Here we use these data to assess the accuracy of the
2dFGRS photometry, the completeness of the parent galaxy catalogue
and the accuracy and reliability of the redshifts.

\subsection{Photometric Accuracy}
\label{sec:photometry} 

The 2dFGRS magnitudes that we use here are the same as those made
public in our June 2001 100k Release.  As described in
\S\ref{sec:sample}, they are pseudo-total magnitudes measured from APM
scans of photographic plates from the UK Schmidt Telescope (UKST)
Southern Sky Survey and their precision depends on the accuracy of the
zeropoint, and non-linearity corrections of each plate, as well as the
measurement errors within each plate.  The overlap of SDSS and 2dFGRS
is in the NGP where our calibration corrections were largest and so
the comparison provides a strong test of these calibrations. 
Note that in the comparisons
made in this section we apply extinction corrections to both the
2dFGRS and SDSS data based on the extinction map of Schlegel \etal
(\shortcite{dust}).

\begin{figure*}%
\epsfxsize=15.5 truecm \epsfbox[-45 -45 658 840]{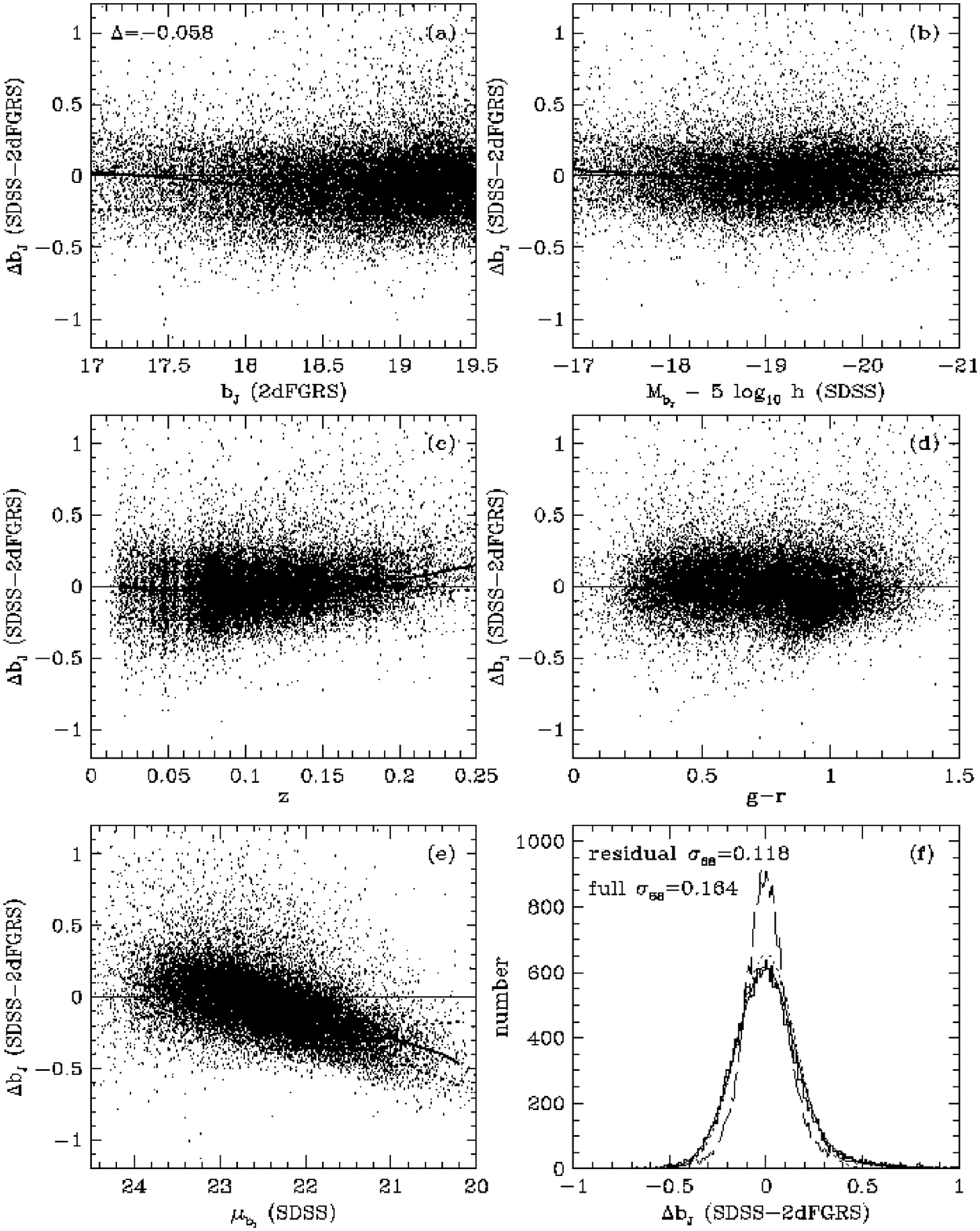}
\caption{Comparison of 2dFGRS photographic $\bj$ magnitudes and CCD
magnitudes from SDSS.  Panel (a) is a scatter plot of the magnitude
difference versus 2dFGRS apparent magnitude and the solid and dotted
lines show the magnitude dependence of the median, 16\% and 84\%
quantiles of the distribution. The median magnitude difference,
$\Delta$, for all the galaxies in the range $17<\bj<19.5$ is indicated
on the panel.  The following four panels show the SDSS-2dFGRS
magnitude differences versus (b) absolute magnitude, (c) redshift,
(d) $\g-\r$ colour and (e) surface brightness.  The surface brightness
is the SDSS measurement of mean surface brightness within the Petrosian half light radius
converted into the $\bj$ band.  
In all these
panels we again show the 16\%, 50\% and 84\% quantiles of the
distribution.  In all but panel (a), the residual is calculated after
subtracting the median offset from each UKST plate.  The distribution
of magnitude differences with respect to the median is shown as a
histogram in panel (f).  The dotted curve, which describes the core of
this distribution quite well, is a gaussian with $\sigma=0.15$
magnitudes.  A robust estimate of the width of this distribution is
$\sigma_{68}=0.164$, where $\pm\sigma_{68}$ spans 68\% of the distribution. 
The empirical model we adopt to describe the full 2dFGRS
magnitude measurement errors is shown by the smooth solid curve in
panel (f) (see text for details). The dashed curve in (f), which
has $\sigma_{68}=0.118$, is the distribution of residual magnitude difference 
relative to the mean correlation of residual with surface brightness
shown in panel (e).
}
\label{fig:mags}
\end{figure*}

The panels of Fig.~\ref{fig:mags} compare 2dFGRS magnitudes
with Petrosian CCD magnitudes from the SDSS EDR (\cite{edr}). 
The definition of the SDSS Petrosian magnitudes is described
in Blanton \etal (\shortcite{blanton}),
which also demonstrates they are essentially equivalent
to total magnitudes for disk galaxies while underestimating the
luminosity of spheroids by approximately 0.15~magnitudes. This
is very similar to the behaviour of the total magnitudes 
derived using Sextractor (\cite{bertin}), which were used in the APM and
2dF calibration. We note that for the purposes of
calibration the less robust SDSS model
magnitudes (the default magnitudes in the SDSS database) 
are not suitable.
Here we have estimated $\bj$ from the SDSS 
photometry\footnote{The calibration of the
magnitudes in SDSS EDR is preliminary. In many of the SDSS papers a
superscript asterix (\eg $\g^*-\r^*$ )
is used to distinguish these magnitudes from those that the SDSS
will ultimately provide. 
}
using the transformation
\begin{equation}
\bj = \g+0.155 +0.152\,(\g-\r).  
\label{eqn:bj}
\end{equation}
This relation comes from adopting the colour equations given for $B$
and $V$ in Fukugita \etal (\shortcite{fukugita96}) and combining these
with $\bj = B -0.28(B-V)$ (Blair \& Gilmore \shortcite{bj}), as 
verified by the EIS data.  Fig.~\ref{fig:mags}d is an empirical test of
the colour term in our adopted transformation. The very weak
dependence of the median magnitude difference on colour is consistent
with the $0.152(\g-\r)$ colour term. A simple least squares fit
gives a coefficient $0.152\pm0.004$ and so is strongly inconsistent with
the colour term of $0.088(\g-\r)$ that was adopted by Blanton \etal 
(\shortcite{blanton})
in the comparison
between the SDSS and the 2dFGRS  luminosity function (\cite{folkes}).

Fig.~\ref{fig:mags}a shows that, in the range $17<\bj<19.5$, the
relation between 2dFGRS and SDSS Petrosian magnitudes is close to
linear and that the scatter between the two measurements is only weakly 
dependent on apparent magnitude, being slightly greater at brighter
magnitudes.  There is evidence for a very weak departure
from linearity with $\Delta \bj^{\rm SDSS} \approx 0.94
\Delta \bj^{\rm 2dF}$ for $\bj<18$, but at fainter magnitudes,
where the vast majority of 2dFGRS galaxies lie,
the relation is accurately linear.
There is a zeropoint offset, with the median 2dFGRS
magnitude being fainter than that of the SDSS by
$\vert\Delta\vert=0.058$~magnitudes.  This is not surprising as the
zeropoint in the SDSS EDR data is only claimed to be accurate to
$\pm0.03$ magnitudes (\cite{blanton}) and similarly we estimate
the accuracy of the 2dFGRS zeropoint to be $\pm0.04$ magnitudes 
(see Section~\ref{100k_cal}).  The SDSS
EDR data span 15 UKST plates in the NGP region of the 2dFGRS and there
is some plate-to-plate variation in the median offset between 2dFGRS
and SDSS Petrosian magnitudes. We find an rms variation of
$0.083$~magnitudes which is in reasonable agreement with the
0.07~magnitudes rms we estimated from the calibrating photometry, and
adds little to the measurement error in an individual galaxy
magnitude. We expect the variation in plate zeropoints to be somewhat
less in the SGP region of the 2dFGRS as this region was constructed
from a more homogeneous set of high quality UKST plates than is
available in the NGP. At present there are not enough public CCD data
to verify this claim. In the other panels of Fig.~\ref{fig:mags} the
median offset between 2dFGRS and SDSS magnitudes on each plate has
been subtracted from the magnitude differences.

Figs.~\ref{fig:mags}b and~c show, for the subset of galaxies for which
redshifts have been measured, the magnitude difference as a function
of absolute magnitude and redshift.  Fig.~\ref{fig:mags}b indicates
that the median offset between 2dFGRS and SDSS magnitudes is, to a good
approximation, independent of absolute magnitude.  In
Fig.~\ref{fig:mags}c below $z \approx 0.16$ there is very little
variation in the median magnitude difference.  At higher redshift
there is a weak trend with the 2dFGRS $\bj$-band magnitude becoming
systematically brighter than that inferred from the SDSS.  We note
that, in contrast, the isophotal magnitudes used by Blanton
\etal (\shortcite{blanton})  in attempting to
reproduce the 2dFGRS LF of Folkes et al. (\shortcite{folkes}), 
which they argued were a good approximation of APM magnitudes, 
falsely predicted that the 2dFGRS
magnitude would monotonically become fainter than the SDSS magnitude
with increasing redshift.  The main reason for the inaccuracy of the
Blanton \etal (\shortcite{blanton}) model is that it neglected to take
account of the way in which APM and 2dFGRS magnitudes are
calibrated. We recall that the calibration of the raw APM magnitudes
involves both a zeropoint and a non-linearity correction so that, in
principle, for galaxies in each interval of apparent magnitude the
median calibrated 2dFGRS magnitude equals the median total magnitude
of the calibrating CCD data (\cite{apmII}). The weak variation with
redshift seen in Fig.~\ref{fig:mags}c is, in fact, probably due to
systematic variation with redshift of the relationship between $\g$,
$\r$ and $\bj$ magnitudes.  The colour equation we have adopted is
empirically verified to be accurate for the bulk of the 2dFGRS
galaxies, which have a median redshift of $z\approx0.1$. At higher
redshift, as different rest frame spectral features pass through the
three filter bands, one expects small changes in the colour equation.

In Fig.~\ref{fig:mags}e we see that there is a significant correlation
between the SDSS-2dFGRS magnitude residual and surface brightness. 
The 2dFGRS magnitudes of galaxies with
surface brightnesses of $\mu_{\bj}\approx23$~mag~arcsec$^{-2}$ are correct
in the mean, but the magnitudes of higher surface brightness galaxies
become progressively too faint.
A similar correlation is also found by
Cross \etal (\shortcite{cross01}) when comparing 2dFGRS and MGC
photometry. Such a correlation is to be expected due to saturation of
the UKST plates (\cite{metcalfe}).  
This variation of measured magnitude with surface
brightness does contribute significantly to the overall 2dFGRS
magnitude measurement error. The distribution shown by the dashed
curve in Fig.~\ref{fig:mags}f shows that correcting for the variation 
with surface brightness would reduce  the width of the error distribution
from $\sigma_{68}=0.164$ to $\sigma_{68}=0.118$
magnitudes.  However, for the full 2dFGRS dataset this cannot be done as
there is not yet a sufficiently accurate measure of surface brightness
(but see Shao \etal \shortcite{apmsb}).  The
surface brightness correlation makes negligible difference to the
overall luminosity function and 2dFGRS selection function.  Here all
one needs is an accurate model of the overall distribution of the
2dFGRS magnitude measurement errors. However if the luminosity
function is derived for subsamples split by spectral type (or another
parameter that correlates with surface brightness) small corrections
have to be made (\cite{madgwick01}).

The histogram in Fig.~\ref{fig:mags}f shows the distribution of
SDSS-2dFGRS magnitude differences.  The dotted curve which
describes the core of the distribution quite well is a gaussian with
$\sigma=0.15$. The tail, in excess of this gaussian, of objects for
which the 2dFGRS measures a fainter magnitude than the SDSS is very
small.  There is a somewhat larger tail of objects for which the
2dFGRS measures a brighter magnitude than the SDSS. It is most likely
that these objects are close pairs of images which the SDSS has
resolved, but which are merged into a single object in the 2dFGRS
catalogue.  This is precisely what is found for the 2dFGRS when
compared to the MGC catalogue (\cite{mgc}) by Cross \etal
(\shortcite{cross01}).  
The overall distribution of magnitude differences
is well fitted by the model shown by the solid curve.  This model is
the sum of a gaussian and a log-normal distribution.  The gaussian
component has $\sigma=0.14$ and accounts for 70\% of the probability
and the remaining 30\% is distributed as a gaussian in
$\ln(1+\Delta\bj)$ with $\sigma=0.235$. We adopt this model as an
empirical description of the distribution of the 2dFGRS magnitude
measurement errors. In so doing, we are assuming that the random
measurement errors in the SDSS CCD Petrosian magnitudes do not
contribute significantly to the width of this distribution. This
assumption is consistent with the comparison of the SDSS photometry
with the deeper MGC CCD photometry in Cross \etal
(\shortcite{cross01}).

\subsection{Completeness of the 2\lowercase{d}F Parent Catalogue}
\label{sec:completeness} 

\begin{figure}
\epsfxsize=8.5 truecm \epsfbox[0 20 530 750]{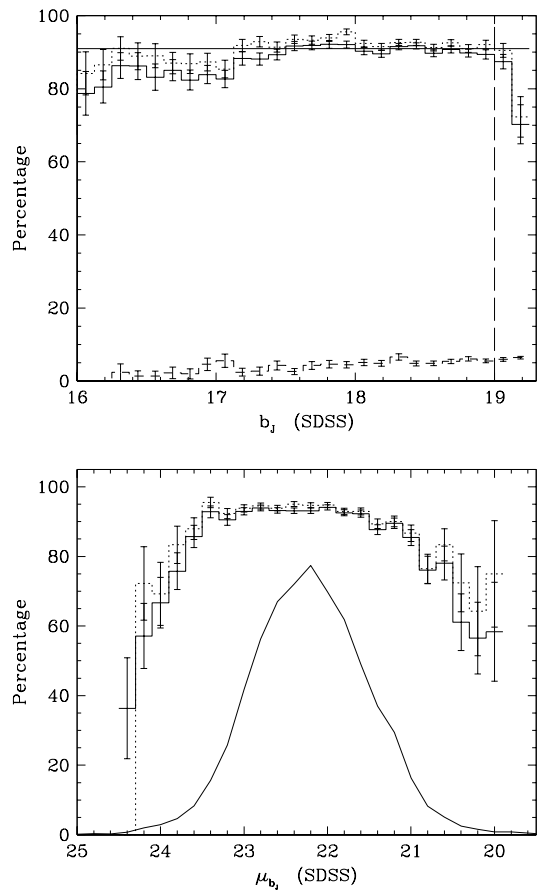}
\caption{In the upper panel the solid histogram shows, as a function 
of apparent magnitude, the percentage of photometrically classified 
galaxies in the SDSS EDR that have 2dFGRS counterparts. This estimate 
of the completeness should be ignored rightwards of the vertical
dashed line at $\bj=19.0$. Fainter than this galaxies
are absent from the 2dFGRS catalogue simply due to the faint
magnitude limit of the catalogue which varies from $19.2<\bj<19.5$.
The horizontal line indicates our adopted 91\% completeness. 
The dotted histogram shows the percentage of 
spectroscopically confirmed SDSS galaxies that have 2dFGRS counterparts. 
The dashed histogram shows the percentage of objects in
the 2dFGRS parent catalogue in the same area that are
spectroscopically identified as stars.  
The lower panel shows the completeness as a function of 
the mean galaxy surface brightness within the Petrosian half light
radius. In this estimate only galaxies in the magnitude range
$17.0<\bj<19.0$ are considered.
Again we show estimates relative to the SDSS 
parent photometric catalogue  (solid histogram)
and the spectroscopically
confirmed galaxies (dotted histogram).
The solid curve shows, on an arbitrary scale,
the distribution of surface brightnesses for the $11\,171$ galaxies
in the SDSS sample in this same magnitude range.
}
\label{fig:class}
\end{figure}

In constructing the parent catalogue of the 2dFGRS the same parameters
and thresholds were used to perform star-galaxy separation as in the
original APM galaxy survey (\cite{apmI}). Thus, the expectation is
that the parent galaxy catalogue will be 90-95\% complete and
contamination from stellar objects will be 5-10\% (\cite{apmI}). In
fact, the spectroscopic identification of the 2dFGRS objects shows
that the stellar contamination is 5.4\% overall and only very weakly
dependent on apparent magnitude (see Fig.~\ref{fig:class}).  The SDSS
EDR allows us to make a useful test of the 2dFGRS galaxy
completeness. In the SDSS commissioning data the star-galaxy
classification procedure is expected to be better than 99\% complete
and have less than 1\% stellar contamination (\cite{blanton}). In
Fig.~\ref{fig:class} we assess the completeness of the 2dFGRS parent
catalogue both against the SDSS photometric galaxy catalogue
and against the SDSS sample of spectroscopically confirmed galaxies.

 To compare to the SDSS spectroscopic sample we selected the 13\,290
SDSS objects that are spectroscopically confirmed as galaxies and have
magnitudes brighter than $\bj=19.5$.  For the SDSS photometric galaxy 
catalogue we used the 16\,371 objects brighter than $\bj=19.5$ that 
are flagged in the EDR database as being members of the SDSS main 
galaxy survey.  The solid (dotted) histogram in the upper panel of 
Fig.~\ref{fig:class} shows, as a function of apparent magnitude,
the percentage of the photometrically classified (spectroscopically confirmed) 
galaxies that have counterparts in the 2dFGRS. 
The completeness estimates vary very little with magnitude over the
entire range $16<\bj<19.0$.  The dip evident in the faintest bins is an 
artifact. Because of random
measurement errors in the APM/2dFGRS magnitudes and because the
magnitude limit in some parts of the  NGP strip is as bright as
$\bj\approx19.2$ (see \cite{colless01} figures~13 and~14), some of the
selected SDSS galaxies have APM magnitudes that are too faint to be
included in the 2dFGRS parent catalogue.  Over the magnitude range
$17.0<\bj<19.0$ the completeness relative to SDSS photometric 
galaxy catalogue is between 88.5\% and 92\%, while the completeness
measured relative to the spectroscopically confirmed SDSS 
galaxy  catalogue is slightly higher at 91\% to 95\%.

The lower panel of Fig.~\ref{fig:class} shows the completeness as a
function of the mean galaxy surface brightness within the Petrosian
half light radius. The surface brightness is estimated from the
extinction-corrected SDSS \g\ and \r\ band data and expressed in the $\bj$
band using the relation given in equation (\ref{eqn:bj}).  In this
estimate only galaxies in the magnitude range $17.0<\bj<19.0$ are
considered. Also shown in this figure is the surface brightness
distribution for galaxies in this magnitude range.  At both extremes
of surface brightness the completeness of the 2dFGRS galaxy catalogue
diminishes. At the faint end this is because the galaxies become to
faint to reliably detect, while at the bright end a fraction are
mis-classified as stars. However over most of surface brightness range
populated by galaxies the completeness of the 2dFGRS is uniformly
high.  Relative to the SDSS catalogue of photometrically classified
galaxies the completeness has a broad peak of 93.4\% for
$21.8<\mu_{\bj}<23.2$, while averaged over the complete galaxy surface
brightness distribution the completeness is 90.9\%. Thus, while the
incompleteness of the 2dFGRS catalogue is approximately 9\%, only
2.5\% incompleteness is dependent on the galaxy surface brightness,
with only one third of this coming from losses at the low surface
brightness end. The dominant reason for the approximately 9\%
incompleteness in the 2dFGRS galaxy catalogue appears to be due to
mis-classification of images.  This conclusion has also been reached
more directly by Pimbblet \etal (\shortcite{pimbblet}),
and Cross \etal (\shortcite{cross01}) by comparing the 2dFGRS parent 
catalogue with deeper wide-area CCD photometry. See also 
Caretta \etal (\shortcite{caretta}), but note the APM catalogue they analyzed
was APMCAT (http://www.ast.cam.ac.uk/$\tilde{\hphantom{i}}$apmcat; 
Irwin \etal \shortcite{apmcatI}; Lewis \& Irwin \shortcite{apmcatII})
which although based on the same scan data as the
Maddox \etal (\shortcite{apmI})
APM galaxy catalogue has a different algorithm for star-galaxy 
classification and less reliable galaxy photometry.
They have shown that while the 2dFGRS
misses some low surface brightness galaxies more are lost due to
mis-classification, particularly of close galaxy and galaxy-star pairs.

The mis-classification of some close galaxy pairs could also explain
the slight difference we have found between the completeness measured
relative to the SDSS spectroscopically confirmed and photometrically
classified galaxy catalogues. Assuming the SDSS spectroscopic sample
is a random sample of the photometric sample, then apart from the
effect of the very small fraction ($<$1\%) of photometrically
classified galaxies which turn out to be stars or artifacts of some
kind, we would expect the two estimates of incompleteness to agree.
However, the spectroscopic sample is not a random sample as close
pairs of galaxies are under-represented because of the mechanical
limits on how close the optical fibres that feed the spectrograph can
be placed.  This is a plausible explanation of the difference between
the two completeness estimates.  We therefore adopt $91\pm2$\% as the
2dFGRS galaxy completeness, consistent with the estimate from the SDSS
photometric catalogue.  This value is indicated by the horizontal line
in the upper panel of Fig.~\ref{fig:class}.

\subsection{Accuracy and Reliability of Redshift Measurements}
\label{sec:redshift} 

\begin{figure}
\epsfxsize=8.5 truecm \epsfbox[80 230 535 725]{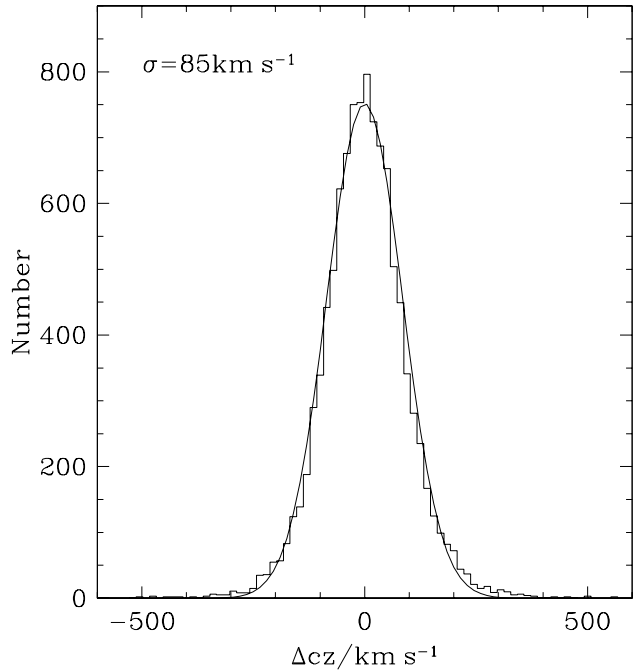}
\caption{
A histogram of the 2dFGRS-SDSS redshift differences for a sample
of 10\,763 galaxies for which both surveys have measured redshifts
with $z>0.003$. The smooth curve is a gaussian with $\sigma=85.0\, \kms$.
}
\label{fig:zhist}
\end{figure}

The 2dFGRS redshift measurements are all assigned a quality flag $Q$
(\cite{colless01}). For most purposes only $Q\ge 3$ redshifts are
used. From a comparison of repeat observations, Colless \etal
(\shortcite{colless01}) estimated that these have a reliability
(defined as the percentage of galaxies whose redshifts are within a
$600\, \kms$ tolerance) of 98.4\% and an rms accuracy of $85\, \kms$.
For higher quality spectra, $Q\ge 4$, these figures improve to $>99$\% and less
than $60\, \kms$, respectively.  Comparison of the 2dFGRS redshifts
with the 10\,763 galaxies which also have redshift measurements in the
SDSS EDR provides a useful check of these numbers. The fraction of
objects for which the redshifts differ by more than $600\, \kms$ is
only 1.0\% and this presumably includes some cases where it is the
SDSS redshift that is in error.
The redshift differences for the remainder are shown in
Fig.~\ref{fig:zhist}. This distribution has a width of $\sigma_{68}=
85.0\, \kms$ (defined so that $2\sigma_{68}$ spans 68\% of the
distribution). Taking account of the contribution from the rms
error in the SDSS measurements this implies a smaller redshift error
than the estimate of Colless \etal (\shortcite{colless01}).  Part of
the reason for the difference in these figures is that the SDSS
galaxies are on average brighter than typical 2dFGRS
galaxies.  Also we have only compared measurements when both the SDSS
and 2dFGRS redshifts are greater than $0.003$. This excludes a small
number of 2dFGRS redshifts that are very small due either to
contamination of the spectra by moonlight or light from a nearby
star. If we further reduce the sample to 10\,022 (or 8\,059) objects by
excluding objects whose SDSS and 2dFGRS positions differ by more than
1 (or 0.5) arc second then the reliability increases slightly to 99.14
(or 99.22\%). This could indicate that some of the discrepant
redshifts arise from very close galaxy pairs that are unresolved in
the 2dFGRS parent catalogue.

\section{\lowercase{k}+\lowercase{e} corrections}
\label{sec:k+e}

\begin{figure}
\epsfxsize=8.5 truecm \epsfbox[20 180 550 625]{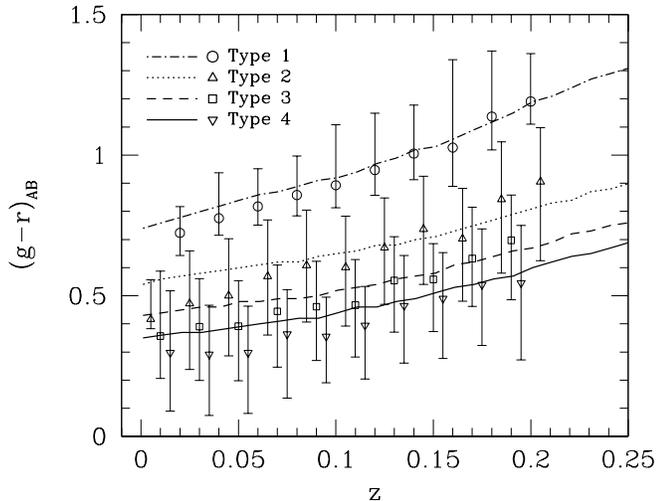}
\caption{Galaxy $\g-\r$ colours as a function of redshift.  The
symbols and error bars show, for each 2dFGRS spectral type, the median,
$10$ and $90$ centiles of the $\g-\r$ colour distribution, as a
function of redshift. The curves are the predictions for model
galaxies, computed using the Bruzual \& Charlot stellar population
synthesis code, whose star formation histories have been selected to
reproduce, as closely as possible, the median colour as a function of
redshift in each class.  }
\label{fig:col_z}
\end{figure}

The final ingredient that is required to characterise the selection
function of the 2dFGRS is a model describing the change in  galaxy magnitude
due to redshifting of the galaxy spectra relative to the $\bj$-filter bandpass 
(k-correction) and galaxy evolution (e-correction). 
These corrections depend on the galaxy's spectrum
and star formation history. As these are correlated, one
can parameterize the k+e corrections as functions of the observed spectra. 

A subset of 2dFGRS spectra have been classified using a method based on
Principal Component Analysis (PCA). A continuous parameter, $\eta$, is
defined as a linear combination of the first two principal components
(\cite{madgwick01}). The definition of $\eta$ is such that its value
correlates with the strength of absorption/emission features.
Galaxies with old stellar populations and strong absorption features
have negative values of $\eta$, while galaxies with young stellar
populations and strong emission lines have positive values.
Therefore, we expect the value of $\eta$ to correlate with the
galaxy's k and k+e correction. In Madgwick \etal
(\shortcite{madgwick01}), the continuous $\eta$ distribution was
divided into four galaxy classes (Type 1: $\eta<-1.4$, Type 2:
$-1.4\le \eta<1.1$, Type 3: $1.1\le\eta<3.5$, Type 4: $3.5\le\eta$) and
the mean k-correction for each type was estimated from the mean
spectra of galaxies in each class.  A current weakness of this
approach is that the overall system response of the 2dF instrument is
not well calibrated. This implies that the resulting k-corrections
have a systematic uncertainty of around 10\% (\cite{madgwick01}).  Due
to this problem and also because we wish to estimate k+e corrections
and not just k-corrections, we have taken a complementary approach.

\begin{figure}
\epsfxsize=8.5 truecm \epsfbox[110 215 575 700]{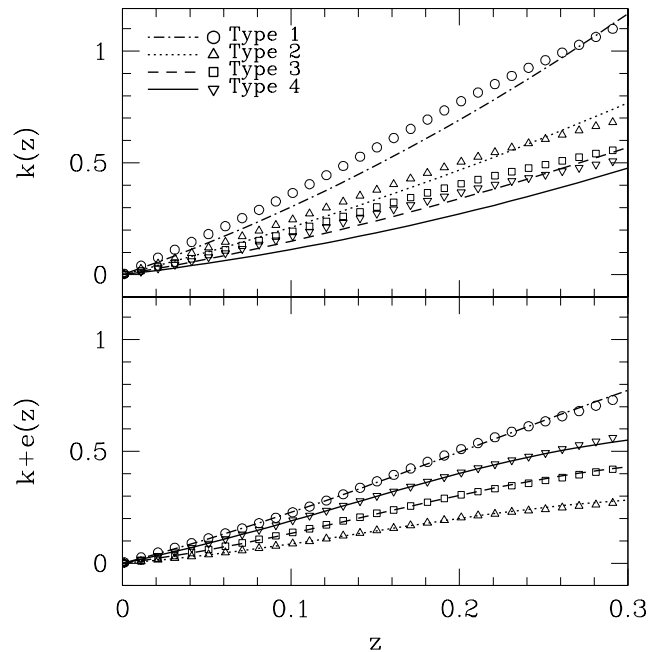}
\caption{Model k and k+e corrections for each 2dFGRS spectral type.
The symbols in the top panel show the k-corrections for four models
selected to match the $\g-\r$ colours as a function of redshift
plotted in Fig.~\ref{fig:col_z}. The curves show the corresponding
k-corrections adopted in Madgwick \etal (2002).   
The symbols in the lower panel show  our model k+e corrections.
In this case, the smooth curves are simple analytic fits
[Type 1 : $k+e = (2z + 2.8  z^2)/(1 +  3.8 z^3)$, Type 2:
$k+e = (0.6 z + 2.8 z^2)/(1 + 19.6 z^3)$, Type 3:
$k+e = (z + 3.6 z^2)/(1 + 16.6 z^3)$, Type 4:
$k+e = (1.6 z + 3.2 z^2)/(1 + 14.6 z^3)$].
}
\label{fig:ke_class}
\end{figure}

The spectrum of any individual galaxy will evolve with time as 
its star formation rate changes and its stellar population evolves. 
Consequently, the spectral type of such a galaxy could vary with 
cosmic time. Therefore, if we want to group the observed
galaxies into discrete classes so that the evolution of
each class can be described by a single model,
we should bin the galaxies in both $\eta$ and $z$. Instead, we
will bin the galaxies only in $\eta$ and so not explicitly
take account of galaxies which evolve from one spectral class
to another. We do this as adopting a more complicated model 
makes little difference to our results and
also it enables us to compare our k-corrections directly with 
those used in Madgwick \etal (\shortcite{madgwick01}).

In Fig.~\ref{fig:col_z}, we plot the median observed $\g-\r$ colour
measured from the SDSS EDR data as a function of redshift for each
spectral class determined from the 2dFGRS spectra.  As expected, we
see that galaxy colour and its dependence on redshift correlates with
the spectral class. Type 1 galaxies, with the most negative value of
$\eta$ and oldest stellar populations, are reddest and Type 4 are
bluest.  The curves plotted on Fig.~\ref{fig:col_z} are models
constructed using the Bruzual \& Charlot (\shortcite{bc93}; in
preparation, see also Liu, Charlot \& Graham \shortcite{lcg} and
Charlot \& Longhetti \shortcite{cl}) stellar population synthesis
code. In a manner very similar to that described by Cole \etal
(\shortcite{irlf}), we ran a grid of models each with the same fixed
metallicity ($Z=Z_\odot/2$) 
and with a star formation history of the form $\psi(t)
\propto \exp(-[t(z)-t(z_{\rm f})]/\tau ) $, with a set of different
timescales, $\tau$. Here, $t(z)$ is the age of the universe at
redshift $z$ and the galaxy is assumed to start forming stars at
$z_{\rm f}=20$. To relate redshift to time, we have assumed a cosmological
model with $\Omega_0=0.3$, $\Lambda_0=0.7$ and Hubble constant $H_0=
70\, \kms$ Mpc$^{-1}$. The k and k+e corrections that we derive are
only very weakly dependent on these choices.

The models plotted Fig.~\ref{fig:col_z} are the four which best
reproduce the observed dependence of the $\g-\r$ colours with redshift
for the four spectral types. They have $\tau=1, 5, 15$ and~$1000$~Gyr
for Type~1,2,3 and~4 respectively. The models provide a complete
description of the galaxy spectral energy distribution and its
evolution and so can be used to define k or k+e corrections for each
spectral type. These are shown by the symbols in
Fig.~\ref{fig:ke_class}.  The Madgwick \etal (\shortcite{madgwick01})
k-corrections, shown by the curves in the top panel, are similar but
systematically smaller than those we have derived. This systematic
difference is comparable to the systematic difference expected given
the current uncertainty in the calibration of the 2dF instrument,
upon which the Madgwick \etal (\shortcite{madgwick01}) k-corrections rely.
The bottom panel of Fig.~\ref{fig:ke_class} shows our k+e
corrections.  Simple analytic fits to the k+e correction for each
spectral class are given in the figure caption and shown by the smooth
curves.  Note that the ordering of the k and k+e corrections is not
the same.  This is because there are competing effects that contribute
to the evolutionary correction. As the redshift increases, the age of
the stellar population viewed decreases. This effect makes galaxies
brighter with increasing redshift, since younger stellar populations
have smaller mass-to-light ratios, and also changes the shape of the
galaxy spectrum. However, there are fewer stars present at earlier
times and this tends to produce a decrease in luminosity with
redshift. For galaxies with ongoing star formation (Types 2,3 and~4)
these effects can all be significant in determining the overall k+e
correction.

\begin{figure}
\epsfxsize=8.5 truecm \epsfbox[30 210 560 660]{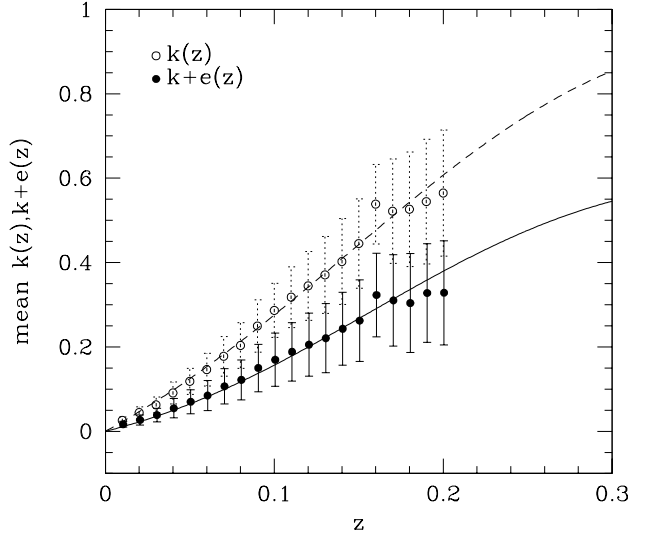}
\caption{The curves show the fits,
$k(z) = (2.2 z + 6 z^2)/(1 + 15 z^3)$ and
$k(z)+e(z) = (z + 6 z^2)/(1 + 20 z^3)$,  to the mean
k and k+e correction as a function of redshift. The mean 
corrections at each redshift, shown by the symbols, have been
computed as a function of redshift 
from the known fractions of each spectral type.
The error bars show the rms scatter about the mean of these distributions. }
\label{fig:ke_mean}
\end{figure}

It is not possible to assign values of $\eta$ to all the galaxies in
the 2dFGRS. In fact, only galaxies with $z<0.2$ are classified in this
way and approximately 5\% of these have spectra with insufficient
signal-to-noise to define $\eta$. Thus, for some purposes it is
necessary to adopt a mean k or k+e correction that can be applied to
all galaxies in the survey.  In Fig.~\ref{fig:ke_mean} we show k and
k+e corrections averaged over the varying mix of galaxies at each
redshift and give simple fitting formulae. We recall that our estimate
of the evolutionary correction assumes a cosmological model with
$\Omega_0=0.3$, $\Lambda_0=0.7$ and $H_0= 70$km\, s$^{-1}$\,Mpc$^{-1}$ in
order to relate redshift and look back time. When estimating the
galaxy luminosity function for cosmological models with different
parameters we retain the same k+e corrections rather than recomputing
the best fitting Bruzual \& Charlot model. While not being entirely
consistent, in practice this makes very little difference to our 
luminosity function estimates. In
section~\ref{sec:lf2df1}, we constrain the uncertainty in k+e
correction by comparing luminosity functions estimated in different
redshift bins. This enables us to assess the contribution to the error
in the luminosity function estimates arising from uncertainties in the
k+e corrections.

\section{Mock and Random Catalogues}
\label{sec:mocks}

One of the main purposes of deriving a quantitative description of the
survey selection function is to make it possible to construct random
(unclustered) and mock (clustered) galaxy catalogues.  The random
catalogues provide a very flexible description of the selection
function and are most often employed when making estimates of galaxy
clustering.  The mock catalogues, where the galaxy positions are
determined from cosmological N-body simulations, are even more useful.
The underlying galaxy clustering and galaxy luminosity function are
known for the mock catalogues and so these catalogues can be
instrumental in testing and developing codes to estimate these
quantities.  They also provide a means for assessing the statistical
errors due to realistic large scale structure on quantities estimated
from the actual redshift survey.  Finally, mock catalogues based on
different cosmological assumptions provide a direct way to compare
clustering statistics for the survey with theoretical predictions.
Here, we briefly describe the steps involved in producing the mock
catalogues that we use below in sections~\ref{sec:lf.norm}
and~\ref{sec:selfun} and that have been employed earlier in other
2dFGRS analysis papers such as Percival \etal (\shortcite{percival}),
Norberg \etal (\shortcite{norbergA,norbergB}). These have been created
from the very large ``Hubble Volume'' simulations carried out by the
Virgo consortium (\cite{evrard99,evrard01}).  
For more details of the
construction of the mock catalogues than are given below see Baugh
\etal (\shortcite{hubblemocks}).

The approach we have taken for generating mock and random catalogues
that match the selection and sampling of the 2dFGRS can be broken into
two stages. In the first stage, we generate idealized mock catalogues,
which have a uniform magnitude limit (somewhat fainter than that of
the true survey) and have no errors in the redshift or magnitude
measurements.  In the second stage, we have the option of introducing
redshift and magnitude measurement errors and we sample the catalogue
to reproduce the slightly varying magnitude limit and the
dependence of the completeness of the redshift catalogue upon
position and apparent magnitude seen in the real 2dFGRS. 
The steps involved in these two
stages are outlined below. In practice, in order to have a fast and
efficient algorithm, some steps are combined, but the result is
entirely equivalent to this simplified description.

\begin{enumerate}
\item{} The first step in generating a mock catalogue consists of
sampling the mass distribution in the N-body simulation so as to
produce a galaxy catalogue with the required clustering. We do this by
applying one of the simple, ad hoc, biasing schemes described by Cole
\etal (\shortcite{oldmocks}). We use their Method 2, but with the
final density field smoothed with a gaussian with smoothing length $
R_S=2 h^{-1}$Mpc and with the parameters $\alpha$ and $\beta$ chosen
to match the observed galaxy power spectrum. For this we took the
galaxy power spectrum of the APM survey (\cite{baugh93}) scaled up in
amplitude by 20\% to match the amplitude of clustering measured in the
2dFGRS at its median redshift.  This results in a fractional rms
fluctuation in the density of galaxies in spheres of $8 h^{-1}$Mpc of
$\sigma_8=0.87$.

\item The second step is to choose the location and orientation of the
observer within the simulation. In the mock catalogues used here, this
was done by applying certain constraints so that the local environment
of the observer resembles that of the Local Group (for details see
Baugh \etal \shortcite{hubblemocks}).

\item We then adopt a Schechter function with
$M_\bj^\star-5\logh=-19.66$, $\alpha=-1.21$ and 
$\Phi^\star=\phistarvaldecon \times 10^{-2} h^3$Mpc$^{-3}$ 
as an accurate description of the present day galaxy luminosity function
(see Section~\ref{sec:lf2df}).
We combine this with the model of the average k+e correction
shown in Fig.~\ref{fig:ke_mean} and the adopted faint survey magnitude
limit to calculate the expected mean comoving space density of galaxies,
$\bar n(z)$, as a function of redshift.

\item We now loop over all the galaxies in the simulation cube that
fall within the angular boundaries of the survey and randomly 
select or reject them so as to produce the required mean $\bar n(z)$.
In the case of random catalogues, we simply generate randomly positioned
points within the boundaries of the survey with spatial number density
given by $\bar n(z)$.

\item For each selected galaxy, we generate an apparent magnitude
consistent with its redshift, the assumed luminosity function and
the faint magnitude limit of the survey. 

\end{enumerate}

To degrade these ideal mock catalogues to match the current
completeness and sampling of the 2dFGRS requires four more steps.

\begin{enumerate}

\item We perturb the galaxy redshifts by drawing random velocities
from a gaussian with $\sigma=85\, \kms$ which is the value
estimated in  Colless \etal (\shortcite{colless01},
see also Section~\ref{sec:redshift}).

\item We perturb the galaxy apparent magnitudes, to account
for measurement errors, by drawing 
random magnitude errors from a distribution that accurately 
fits the histogram of SDSS-2dFGRS magnitude differences shown
in Fig.~\ref{fig:mags}f.

\item We make use of the map of the survey magnitude limit
as a function of position to throw out galaxies that would be
too faint to have been included in the actual 2dFGRS parent catalogue.

\item The final step incorporates the current level of
completeness of the 2dFGRS redshift catalogue.  Here, we make use of
the maps $R(\btheta)$ and $S(\btheta,\bj)$, which quantify the
completeness of the survey. They are defined in
Section~8 of Colless \etal (\shortcite{colless01}) and summarised
in Appendix~\ref{app:maps}. At each angular position, $\btheta$, 
only a fraction,
$R(\btheta)$, of the redshifts is retained or, taking account of the slight
dependence of completeness upon the apparent magnitude, 
a fraction $S(\btheta,\bj)$, which depends upon apparent
magnitude, $\bj$, as well as position, is instead retained.

\end{enumerate}

\section{The 2\lowercase{d}FGRS luminosity function for different sub-samples}
\label{sec:lf}

The luminosity functions presented here are estimated using
fairly standard implementations of the STY (Sandage, Tammann \& Yahil 
\shortcite{sty}) and stepwise maximum likelihood (SWML 
Efstathiou, Ellis \& Peterson \shortcite{swml}) estimators.
The only modifications we have made to the methods described in these
papers are:
\begin{enumerate}
\item We use the map, $\bj^{\rm lim}(\btheta)$, of the survey magnitude
limit to define the apparent magnitude limit for each individual galaxy.
\item We use the map of $\mu(\btheta)$ to define a weight,
$1/c_z(\bj,\mu[\btheta])$, for each galaxy (see equation \ref{eqn:compl})
to compensate for the magnitude dependent incompleteness.
\end{enumerate}
Provided the most incomplete 2dF fields are excluded from the sample,
then the variation in these weights is small. Slightly more than 76\%
of the observed 2dF fields have an overall redshift completeness
greater than 90\%. Here we exclude the few fields for which the redshift
completeness is below 70\%. For this sample the mean weight is
$1.06$ and the rms variation about this is only $0.06$.  Furthermore, one
can make the influence of the weight completely negligible by applying
an additional magnitude cut and discarding galaxies fainter than, for
example, $\bj=19.2$.

We have applied both our STY and SWML LF estimators to galaxy samples
extracted from the mock galaxy catalogues. In the case of the idealized
mock catalogues, not only do the mean estimated luminosity functions
agree precisely with the input luminosity function, but also the
error estimates agree well with the scatter between the estimates
from the 22 different mock catalogues.  
For the degraded mocks the estimated luminosity functions reproduce
well the input luminosity functions convolved with the assumed magnitude
errors. It is perhaps also worth noting that we checked that 
the independently written STY code used in Madgwick \etal 
(\shortcite{madgwick01})  gave identical results when applied to the same 
sample and assuming the same k-corrections.

\begin{figure*}
\epsfxsize=17.5 truecm \epsfbox[17 195 530 750]{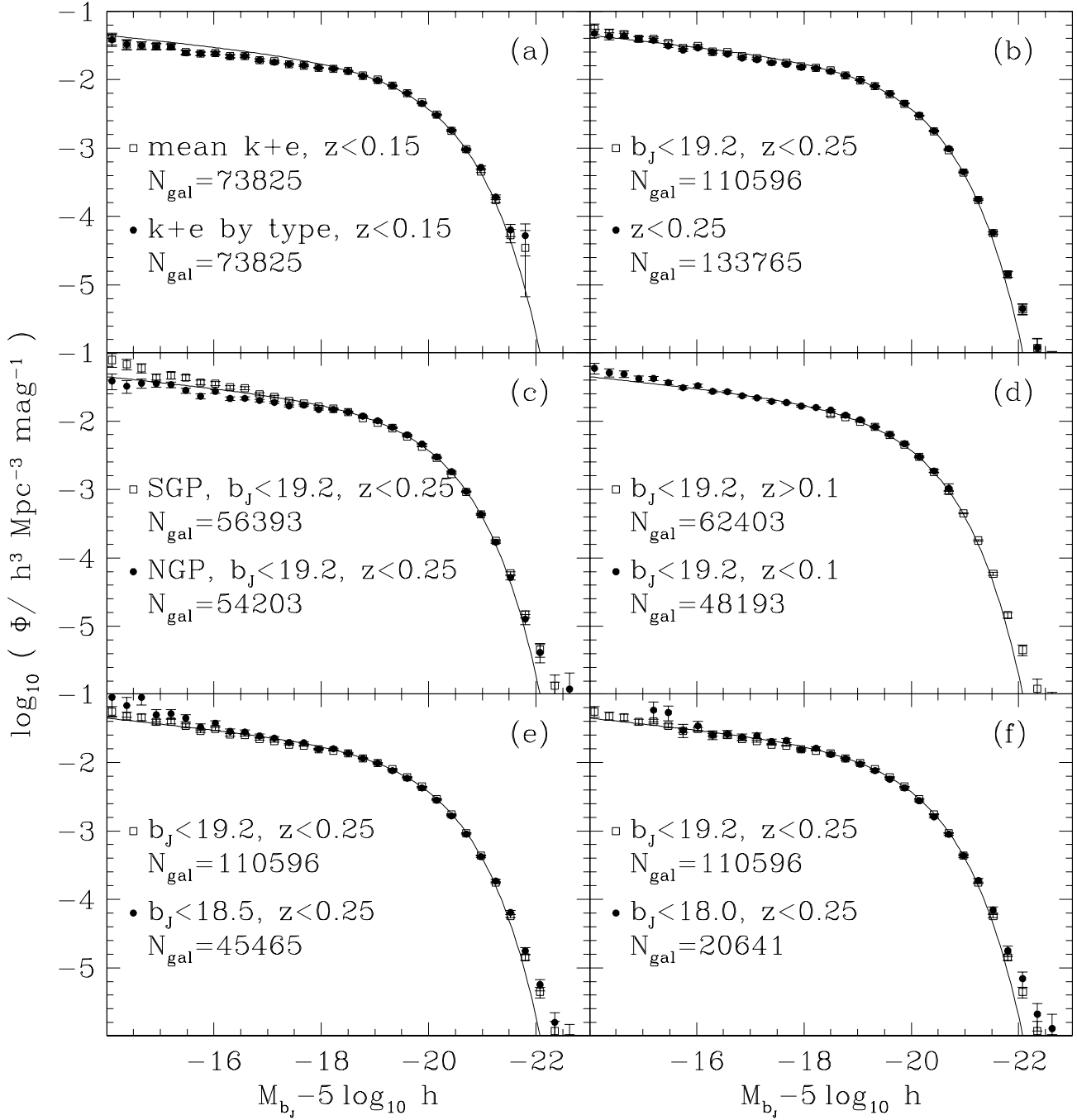}
\caption{Luminosity functions for different subsamples of the 2dFGRS
data.  The smooth curve in each panel is a Schechter function with
$M_\bj^\star-5\logh=-19.67$, $\alpha=-1.21$ and $\Phi^\star=\phistarval\times
10^{-2}h^3$Mpc$^{-3}$.  This is the STY estimate for the sample defined
by $17<\bj<19.2$ and $z<0.25$ and computed using the average k+e
correction shown in Fig.~\ref{fig:ke_mean}.  This curve is reproduced
in each panel as a fiducial reference.  In the different panels the points 
and error bars show SWML LF estimates for different
subsets of the 2dFGRS as indicated by the selection criteria given in
each legend (see text for details).  Also indicated on each panel is
the number of galaxies in each sample.  An $\Omega_0=0.3$,
$\Lambda_0=0.7$ cosmology is assumed and the luminosity functions have
been normalized to produce 146 galaxies per square degree brighter
than $\bj=19.2$.  }
\label{fig:lf6}
\end{figure*}

Due to the large size of the 2dFGRS the statistical errors in our
estimated luminosity functions are extremely small.  It is therefore
important to verify that systematic errors are well controlled.  This
is partially demonstrated in Fig.~\ref{fig:lf6}, where we compare
LF estimates for various subsamples of the 2dFGRS.

For all the samples shown in Fig.~\ref{fig:lf6} we have applied a
bright magnitude cut of $\bj>17$ and assumed an $\Omega_0=0.3$,
$\Lambda_0=0.7$ cosmology.  In addition, we have applied various extra
cuts to define different subsamples. The smooth curve in each panel of
Fig.~\ref{fig:lf6} is a Schechter (\shortcite{schechter}) function,
\begin{equation} 
{d\Phi \over dM} = 0.921 \, \Phi^* \, (L/L^\star)^{\alpha+1} \, \exp(-L/L^\star)
\end{equation} 
where the magnitude corresponding to the luminosity $L^\star$
is $M_\bj^\star-5\logh=-19.67$, $\alpha=-1.21$ and
$\Phi^\star=\phistarval \times 10^{-2}$Mpc$^{-3}$.  This is the STY
estimate for the sample defined by $17<\bj<19.2$ and $z<0.25$.  In
both the STY and SWML LF estimates, the normalization of the
luminosity function is arbitrary. To aid in the comparisons shown in
Fig.~\ref{fig:lf6}, we have normalized each estimate to produce 146
galaxies per square degree brighter than $\bj=19.2$ (see
Section~\ref{sec:lf.norm}).  It can seen by comparison with the SWML
estimates in each panel that the Schechter function is not a good fit
at the very bright end. However, it should be borne in mind that in
these estimates we have made no attempt to correct for the magnitude
measurement errors.  Thus, these luminosity functions all represent
the true luminosity function convolved with the magnitude measurement
errors.

The influence of the assumed k+e correction is investigated in
Fig.~\ref{fig:lf6}a. The sample used for both the estimates in this
panel is defined by the limits $17<\bj<19.2$ and $z<0.15$ and includes
only galaxies which have been assigned a spectral type.
The upper redshift limit is imposed to
avoid the interval where contamination by sky lines causes the
spectral classification to be unreliable (\cite{madgwick01}). 
For one sample, we use the average k+e
correction shown in Fig.~\ref{fig:ke_mean}, while for the other, we
adopt the spectral class dependent k+e corrections of
Fig.~\ref{fig:ke_class}. We see that the two estimates agree
very accurately at all magnitudes. As the systematic difference 
is so small, we adopt for all other estimates the
global k+e correction which then allows us to use the full 
redshift sample. The samples analysed in this panel exclude 
a small fraction  (5\%) of galaxies whose spectra have 
insufficient signal-to-noise to enable
spectral classification. These are typically low surface brightness,
low luminosity galaxies. It is for this reason that the luminosity
function estimates in this panel fall slightly below the estimates
in the other panels for magnitudes fainter than $M_\bj-5\logh=-17$.

Fig.~\ref{fig:lf6}b shows SWML estimates for samples including
galaxies with redshifts up to $z=0.25$. The two estimates compare the
results for a sample limited by $\bj<19.2$ and the sample to the full
depth of the 2dFGRS, which has a spatially varying magnitude limit of
$19.4\pm0.1$ (see figures~13 and~14 of Colless \etal
\shortcite{colless01}). The close agreement between the two indicates
that no significant bias or error has been introduced by taking
account of the varying magnitude limit and including the correction
for the magnitude dependent incompleteness.

The remaining panels of Fig.~\ref{fig:lf6} all use samples limited
by $\bj<19.2$, but essentially identical results are found if the samples
are extended to the full depth of the survey. Fig.~\ref{fig:lf6}c compares
the LF estimates from the spatially separated SGP and
NGP regions of the 2dFGRS. Brighter than $M_\bj^\star-5\logh=-17.5$,  
the two regions yield luminosity functions with identical shapes.
Note that both luminosity functions have been normalized to produce
146 galaxies per square degree brighter than
$\bj=19.2$, rather than to the actual galaxy number counts in each region. 
This good agreement suggests that any systematic offset in zeropoint of
the magnitude scale in the two disjoint regions is very small.
If one allows an offset between the zeropoints of the NGP and
SGP magnitude scales, then comparing the bright ends of
these two luminosity functions ($M_\bj-5\logh<-17.5$)  
constrains this offset to the rather small value  $0.014\pm0.01$.
Fainter than $M_\bj-5\logh=-17.5$  the two estimates differ 
systematically to a small but significant degree.
We return to this difference briefly in Section~\ref{sec:lf2df1}.

Fig.~\ref{fig:lf6}d compares results from samples split by redshift.
Here, the combined effect of the redshift and apparent magnitude
limits results in estimates that only span a limited range in absolute
magnitude. To normalize these luminosity functions we extrapolated the
estimates using their corresponding STY Schechter function estimates.
The two luminosity functions agree well in the overlapping magnitude range
and also agree well with the full samples shown in the other panels.
This demonstrates that the evolution of the luminosity function is consistent
with the k+e-correction model we have adopted. Since we apply k+e 
corrections, the luminosity function we estimate is always that at $z=0$.

The final two panels in Fig.~\ref{fig:lf6} examine  luminosity functions
estimated from bright subsamples of the 2dFGRS. Fig.~\ref{fig:lf6}e
shows an estimate for galaxies brighter than $\bj=18.5$ and
Fig.~\ref{fig:lf6}f for galaxies brighter than $\bj=18.0$. The
statistical errors in the estimates from these smaller samples are
significantly larger. Nevertheless, the luminosity functions agree
well, on average, with those from the deeper samples.

\section{Galaxy Number Counts}
\label{sec:lf.norm}

\begin{figure}
\epsfxsize=8.5 truecm \epsfbox[0 60 365 750]{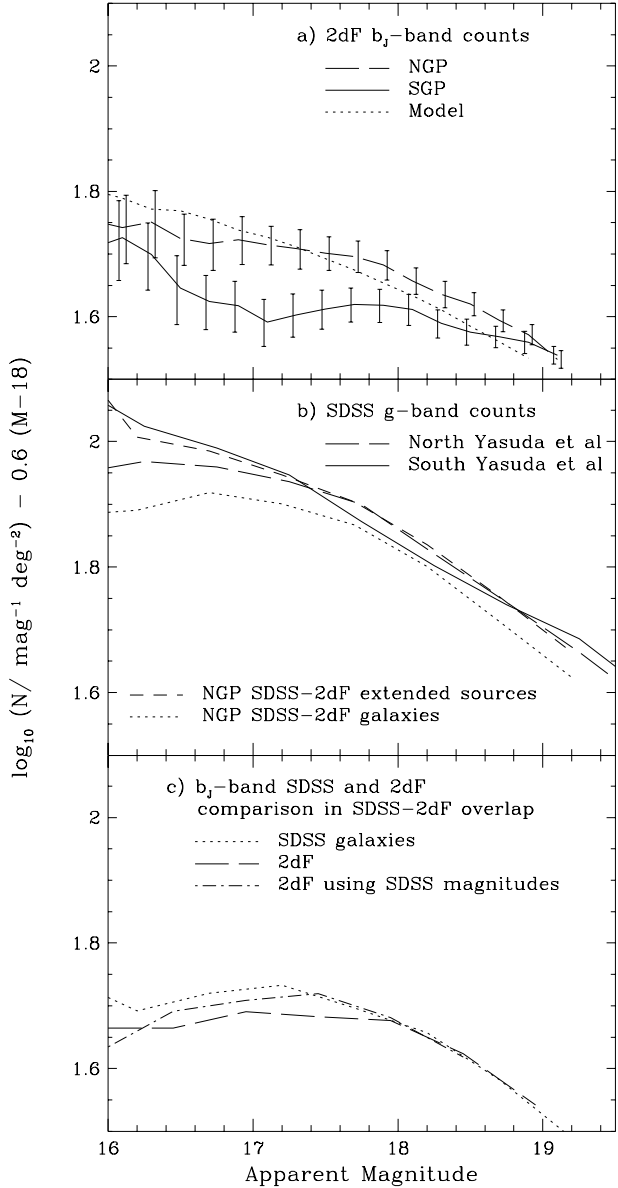}
\caption{The 2dFGRS and SDSS galaxy number counts in the
$\bj$ and $\g$-bands. In each panel we plot the logarithm of the number
of galaxies per unit apparent magnitude scaled by a
Euclidean model. This enables the ordinate to be expanded so that
small differences in the counts are visible. The upper panel shows
the 2dFGRS $\bj$-band counts separately in the NGP and SGP regions.
The error bars show the rms variation we expect due to large scale
structure, estimated from our 22 mock catalogues.
The middle panel compares the published SDSS $\g$-band counts 
of Yasuda \etal (2001) and our own estimates of the SDSS counts in the area 
which overlaps with the 2dFGRS NGP region. The bottom panel compares, 
in the overlap region, SDSS and 2dFGRS $\bj$-band counts.
}
\label{fig:euclid}
\end{figure}

In the previous section we have demonstrated  that the
shape of the 2dF galaxy luminosity function, brighter
than $M_\bj-5\logh<-17$, is robust to
variations in the sample selection and the assumed k+e corrections.
We have not yet addressed the issue of normalization and its
uncertainty; we simply normalized all the estimates to produce 146 galaxies
per square degree brighter than $\bj=19.2$. 
We now investigate the uncertainty in this normalization due to 
both large scale structure and the uncertainty in systematic corrections.

\subsection{The 2\lowercase{d}FGRS b$\bf _J$-band galaxy counts}

The upper panel in Fig.~\ref{fig:euclid} shows the 2dFGRS galaxy
$\bj$-band number counts in the NGP and SGP. In this figure we have
subtracted a Euclidean model from the counts to enable the ordinate to
be expanded so that small differences are visible. These are counts of
objects in the 2dFGRS parent catalogue (after the removal of the
merged images that did not form part of the 2dFGRS target list)
multiplied by a factor of $1/(1.054\times 0.91)=1.043$ to take account of the
stellar contamination (5.4\%) and incompleteness (9\%) discussed in
Section~\ref{sec:completeness}. While these numbers are derived from a
comparison with the SDSS EDR we note that they are very close to
the original estimates given by Maddox \etal (\shortcite{apmI}).  The
error bars placed on the measured counts are the rms scatter seen in
our 22 mock catalogues and provide an estimate of the variation
expected due to large scale structure. The dotted curve is the mean
number counts in the mocks and corresponds to the expectation for a
homogeneous universe.

It has long been known that the galaxy counts in the APM catalogue are
steeper than model predictions for a homogeneous universe
(\cite{apm_counts}). As we have subtracted the Euclidean slope this
manifests itself in Fig.~\ref{fig:euclid} as a shallower slope for the
SGP curve than the model prediction shown by the dotted curve.  
The model assumes $\Omega_0=0.3$, $\Lambda_0=0.7$,
the luminosity function estimated in the previous
section and the mean k+e-correction estimated in Section~\ref{sec:k+e}.
The NGP counts are greater than those in the SGP throughout the range
$16<\bj<19$ and are also slightly steeper than the model prediction
(i.e. shallower in Fig.~\ref{fig:euclid}), although the difference is
not as extreme as for the SGP.  The 1-$\sigma$ error bars determined
from the mock catalogues show that deviations from the homogeneous
model prediction such as those shown by the NGP should be common. The
SGP counts are harder to reconcile with the model, but it should be
borne in mind that even on quite large scales the galaxy density field
is non-gaussian and so 1-$\sigma$ error bars do not fully quantify the
expected variation.

To normalize our estimates of the galaxy luminosity function we use
the cumulative count of galaxies per square degree brighter than
$\bj=19.2$. In the $740\,$deg$^2$ of the NGP strip this is $ 151.6\pm
6.1 $, where the error is again the rms from the mock catalogues. The
corresponding numbers for the $1094\,$deg$^2$ SGP strip are $ 141.4\pm
6.1 $ and, for the combined $1834\,$deg$^2$, $146\pm 4.4 $.  The NGP
and SGP number counts differ by 7\%, but this is reasonably common in
the mock catalogues.

\subsection{Comparison of 2\lowercase{d}FGRS and SDSS counts}

 The middle panel in Fig.~\ref{fig:euclid} shows SDSS $\g$-band counts
(this being the SDSS band closest to $\bj$). We show both the
published SDSS counts from Yasuda \etal (\shortcite{yasuda}) and
two estimates we have made directly from the SDSS EDR that overlaps with
2dFGRS NGP strip. The counts shown by the short dashed curve are of
extended sources that satisfy the criterion used in
Yasuda \etal (\shortcite{yasuda}) of
$\r_{\rm PSF} - \r_{\rm model} >0.145$. 
This criterion, which compares an estimate of the
magnitude of an object assuming it to be a point source with an
estimate obtained by fitting a model galaxy template, is very
effective at rejecting faint stars from the sample. 
The very accurate agreement between the
published northern counts and our estimate from the EDR data
demonstrates that the simple star-galaxy classification criterion we
have used works well fainter than $\g=17.0$ and that we have correctly
estimated the area of the overlap between the SDSS EDR and the NGP
region of the 2dFGRS. The Yasuda \etal (\shortcite{yasuda}) counts are
accurate brighter than $\g=17.0$ as at brighter magnitudes
they utilise a more sophisticated star-galaxy separation algorithm 
supplemented by visual classification. The galaxy counts shown by the
dotted curve are the counts of objects in the EDR database which meet
all the criteria, excluding the cut on \r-band magnitude, for inclusion
in the SDSS main galaxy survey.  These counts are systematically
8.4\% lower than our estimate of the SDSS extended source counts 
and also the galaxy counts of Yasuda \etal (\shortcite{yasuda}). 
The reason for this is that for inclusion in the SDSS main galaxy survey 
the sources have to satisfy additional criteria described in 
Strauss \etal (\shortcite{strauss02}). First 
a stricter extended source criterion ($\r_{\rm PSF} - \r_{\rm model} >0.3$)
rejects an additional 2.1\% of the objects. A surface brightness
threshold of $\mu_\r<24.5$ (comparable to $\mu_\bj=25.6$) rejects
a further 4.1\%. Rejection of images containing saturated pixels 
(probably stars) removes a further 1.6\% and lastly 0.6\% of images
are rejected as blended. Strauss \etal (\shortcite{strauss02})
conclude that the galaxy sample they define has a completeness
exceeding 99\%. This is consistent with the very small (0.36\%)
of spectroscopically confirmed 2dFGRS galaxies whose counterparts
in the SDSS survey do not satisfy the Strauss \etal (\shortcite{strauss02}) 
galaxy selection criteria. We, therefore, conclude the that the
Yasuda \etal counts are biased high.

The lower panel of Fig.~\ref{fig:euclid} compares SDSS and 2dFGRS
$\bj$ galaxy counts within the approximately $173\,$deg${^2}$ area of overlap
of the two datasets.  Note that this is essentially the whole of the
northern SDSS data. Only small areas are discarded where satellite trails
and other defects have been cut out of the 2dFGRS sky coverage.
Here, we have estimated $\bj$ from the SDSS
Petrosian magnitudes using equation \ref{eqn:bj}, but also including
explicitly the $0.058$ magnitude zeropoint offset we measured in
Section~\ref{sec:photometry}.  We see that between $18<\bj<19$, the
2dFGRS and SDSS number counts agree very accurately. In this area the
cumulative count of galaxies per square degree brighter than
$\bj=19.2$ is $150$, 5\% higher than the average over the ten times
larger area covered by the combined NGP+SGP 2dFGRS strips.  Between
$17<\bj<18$ the 2dFGRS counts are approximately 8\% below the SDSS
counts.  This accounted for by the slight non-linearity 
we noted in Section~\ref{sec:photometry} 
between the bright ($\bj<18$) SDSS and 2dFGRS magnitudes.
If we compute the counts for the same
2dFGRS objects, but using the magnitudes derived from the SDSS data
then there is better agreement between 2dFGRS and SDSS for
$17<\bj<18$. Brighter than $\bj=17$ the slight decrease in completeness
of the 2dFGRS catalogue evident in Fig.~\ref{fig:class}
also contributes to a modest reduction in the 2dFGRS galaxy counts.

We conclude from this comparison that in the $173\,$deg${^2}$ region of
overlap, the 2dFGRS counts (corrected using the standard estimates of 
stellar contamination and incompleteness) are in good agreement with
the SDSS galaxy counts fainter than $\bj=17$, but are 5\% higher than those
averaged over the full area of the 2dFGRS. The 1-$\sigma$ statistical
error estimated from the mock catalogues for an area this size is
4.8\%. Over the full area,
we find $146$ galaxies per square degree brighter than $\bj=19.2$
with a 1-$\sigma$ statistical error, estimated from mock catalogues, 
of just 3\%.

\section{The Normalized 2\lowercase{d}FGRS Luminosity Function }
\label{sec:lf2df}

We now use the number counts to normalize our LF estimates. 
In doing this we integrate the estimated LF over the absolute magnitude
range $-13>\Mag>-24$. The contribution to the counts from galaxies
outside this range is completely negligible.

\subsection{Independent NGP and SGP Estimates}
\label{sec:lf2df1}

\begin{figure*}
\epsfxsize=14.5 truecm \epsfbox[20 40 530 745]{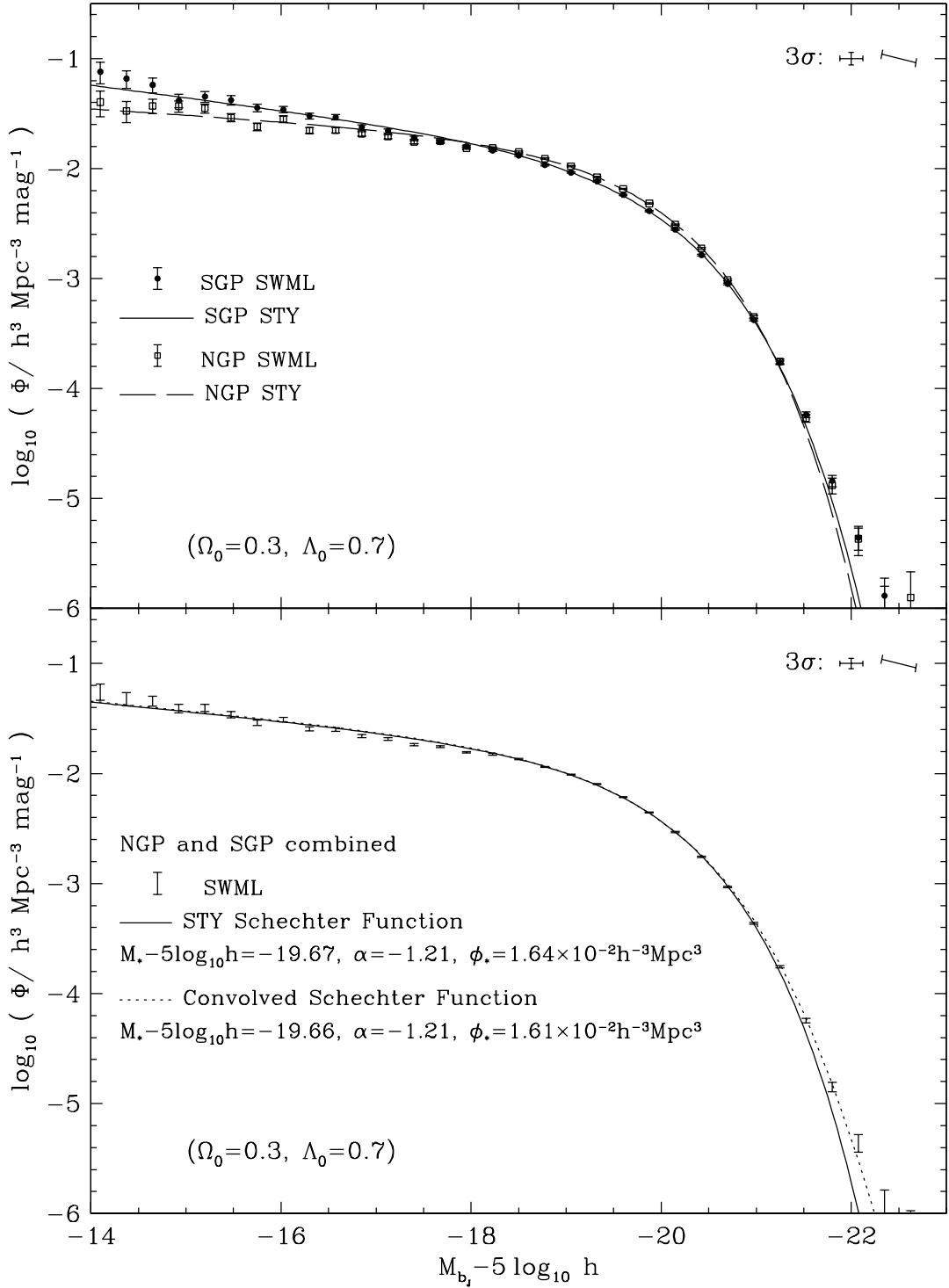}
\caption{ The upper panel shows two independent estimates of the
$z=0$ galaxy luminosity function, from the NGP and SGP regions.
The lower panel shows the combined NGP+SGP estimate, 
normalized to the mean NGP+SGP number counts.
The symbols show SWML estimates with $\pm1\sigma$ error bars
and the smooth solid curves STY
Schechter function estimates. The dotted curve in the lower 
panel is the fit to the SWML LF obtained using a Schechter function convolved
with the distribution of magnitude measurement errors.
The parameters of the Schechter functions are given in the legend.
The error bars shown in the upper right of each panel are 
$3\sigma$ (for clarity) errors showing the additional uncertainty
in the normalization (vertical), in the photometric zeropoint
(horizontal) and in the k+e-corrections (slanted).
These three sources of error are all independent, but affect
each data point in the luminosity function coherently.
Here, and in all our plots, an $\Omega_0=0.3$ and $\Lambda_0=0.7$ 
cosmology is assumed. The values of the SWML estimate are
given in Table~\ref{tab:swml} and the parameters of the 
deconvolved Schechter function fits are given
in Table~\ref{tab:sty}, along with estimates for alternative
choices of the cosmological parameters.
}
\label{fig:lf_main}
\end{figure*}

In the upper panel of Fig.~\ref{fig:lf_main} we present two
independent estimates of the galaxy luminosity function, from the NGP
and SGP regions. Here, the LF estimate in each region is normalized by
its own galaxy number counts. Thus, the two estimates are independent
and the differences between them provide an estimate of the
statistical errors. These can be compared with the plotted SWML
errors, but note should be taken that the SWML errors do not take
account of the uncertainty in the normalization of the luminosity function.
 For these
two estimates, the mock catalogues indicate that the contribution to
the uncertainty of the normalization 
from large scale structure is about 4\%.  Also of
importance is the uncertainty in the incompleteness corrections.  We
have corrected assuming a global 9\% incompleteness in the 2dFGRS
photometric catalogue and the uncertainty in this adds, in quadrature,
approximately 2\% to the normalization uncertainty (see
Section~\ref{sec:photometry}).  An indication of this uncertainty is
given by the vertical error bar plotted in the upper right of each
panel of Fig.\ref{fig:lf_main}, which, for clarity, shows the
$\pm3\sigma$ range.  If this is added in quadrature to the SWML
errors, then one finds that the differences between the NGP and SGP
estimates are entirely consistent except for magnitudes fainter than
$\Mag=-17.5$.

At the faint end, the SGP LF is slightly steeper than that estimated
from the NGP.  This may reflect genuine spatial variations in the
galaxy luminosity function as this faint portion of the luminosity
function is determined from a very local volume.  Such variations are
perhaps to be expected given the results of Norberg \etal
(\shortcite{norbergA,norbergB}) that show that galaxies of different
luminosity have systematically different clustering properties. The
faint end of the luminosity function may also be affected by
incompleteness and magnitude errors in the 2dFGRS.  We have corrected
the luminosity function assuming that the incompleteness and magnitude
errors are independent of absolute magnitude. However, from the joint
analysis of the 2dFGRS and the much deeper MGC catalogue by Cross
\etal (\shortcite{cross01}), we know that the magnitude errors are
largest for objects of extreme surface brightness and also part of the
incompleteness is due to the 2dFGRS preferentially missing very low
surface brightness galaxies.  The correlation between absolute
magnitude and surface brightness (\cite{sbI,sbII}) then implies that
low luminosity galaxies are underrepresented.  The work of Cross \&
Driver (\shortcite{cd01}) (see also Cross \etal \shortcite{cross01})
suggests that this only becomes important fainter than $\Mag=-16.5$.

\begin{table*}
\caption{The stepwise maximum likelihood (SWML) estimates of the 
2dFGRS $z=0$ galaxy luminosity function for three assumed cosmological
 models. The quoted errors do not take account of uncertainty in the 
normalization, the photometric zeropoint or uncertainty in the appropriate 
evolutionary correction (see Section~\ref{sec:lf.norm}). Also these 
estimates are not deconvolved for the effect of random
magnitude measurement errors.
}
\begin{center}
\begin{tabular}{llrrrrrrrrrrrrrrr} 
\multicolumn{1}{l} {} &
\multicolumn{1}{l} {$\Omega_0=0.3$, $\Lambda_0=0.7$} &
\multicolumn{1}{l} {$\Omega_0=1$, $\Lambda_0=0$} &
\multicolumn{1}{l} {$\Omega_0=0.3$, $\Lambda_0=0$}  \\
\multicolumn{1}{l} {M$_\bj - 5\logh$} &
\multicolumn{1}{l} {$\Phi$/$h^3$Mpc$^{-3}$mag$^{-1}$} & 
\multicolumn{1}{l} {$\Phi$/$h^3$Mpc$^{-3}$mag$^{-1}$} & 
\multicolumn{1}{l} {$\Phi$/$h^3$Mpc$^{-3}$mag$^{-1}$}  \\
\hline 
$ -13.275 $ & $ (8.850 \pm 4.560) \times 10^{-2} $ & $ (1.100 \pm 0.565) \times 10^{-1} $ & $ (1.094 \pm 0.563) \times 10^{-1} $ \\ 
$ -13.550 $ & $ (5.344 \pm 1.839) \times 10^{-2} $ & $ (6.283 \pm 2.147) \times 10^{-2} $ & $ (6.126 \pm 2.096) \times 10^{-2} $ \\ 
$ -13.825 $ & $ (5.642 \pm 1.289) \times 10^{-2} $ & $ (6.438 \pm 1.463) \times 10^{-2} $ & $ (6.222 \pm 1.415) \times 10^{-2} $ \\ 
$ -14.100 $ & $ (5.580 \pm 0.957) \times 10^{-2} $ & $ (6.145 \pm 1.058) \times 10^{-2} $ & $ (5.503 \pm 0.977) \times 10^{-2} $ \\ 
$ -14.375 $ & $ (4.785 \pm 0.661) \times 10^{-2} $ & $ (5.535 \pm 0.751) \times 10^{-2} $ & $ (5.453 \pm 0.737) \times 10^{-2} $ \\ 
$ -14.650 $ & $ (4.573 \pm 0.487) \times 10^{-2} $ & $ (5.140 \pm 0.540) \times 10^{-2} $ & $ (4.922 \pm 0.522) \times 10^{-2} $ \\ 
$ -14.925 $ & $ (3.927 \pm 0.351) \times 10^{-2} $ & $ (4.398 \pm 0.388) \times 10^{-2} $ & $ (4.300 \pm 0.379) \times 10^{-2} $ \\ 
$ -15.200 $ & $ (3.963 \pm 0.288) \times 10^{-2} $ & $ (4.558 \pm 0.324) \times 10^{-2} $ & $ (4.318 \pm 0.311) \times 10^{-2} $ \\ 
$ -15.475 $ & $ (3.437 \pm 0.222) \times 10^{-2} $ & $ (3.889 \pm 0.246) \times 10^{-2} $ & $ (3.746 \pm 0.239) \times 10^{-2} $ \\ 
$ -15.750 $ & $ (2.906 \pm 0.166) \times 10^{-2} $ & $ (3.304 \pm 0.184) \times 10^{-2} $ & $ (3.186 \pm 0.179) \times 10^{-2} $ \\ 
$ -16.025 $ & $ (3.096 \pm 0.148) \times 10^{-2} $ & $ (3.517 \pm 0.164) \times 10^{-2} $ & $ (3.383 \pm 0.159) \times 10^{-2} $ \\ 
$ -16.300 $ & $ (2.555 \pm 0.111) \times 10^{-2} $ & $ (2.940 \pm 0.124) \times 10^{-2} $ & $ (2.800 \pm 0.120) \times 10^{-2} $ \\ 
$ -16.575 $ & $ (2.522 \pm 0.098) \times 10^{-2} $ & $ (2.871 \pm 0.108) \times 10^{-2} $ & $ (2.735 \pm 0.104) \times 10^{-2} $ \\ 
$ -16.850 $ & $ (2.198 \pm 0.075) \times 10^{-2} $ & $ (2.532 \pm 0.082) \times 10^{-2} $ & $ (2.429 \pm 0.080) \times 10^{-2} $ \\ 
$ -17.125 $ & $ (2.055 \pm 0.059) \times 10^{-2} $ & $ (2.365 \pm 0.064) \times 10^{-2} $ & $ (2.283 \pm 0.063) \times 10^{-2} $ \\ 
$ -17.400 $ & $ (1.826 \pm 0.043) \times 10^{-2} $ & $ (2.043 \pm 0.046) \times 10^{-2} $ & $ (1.961 \pm 0.045) \times 10^{-2} $ \\ 
$ -17.675 $ & $ (1.757 \pm 0.035) \times 10^{-2} $ & $ (1.996 \pm 0.038) \times 10^{-2} $ & $ (1.911 \pm 0.037) \times 10^{-2} $ \\ 
$ -17.950 $ & $ (1.560 \pm 0.027) \times 10^{-2} $ & $ (1.812 \pm 0.029) \times 10^{-2} $ & $ (1.736 \pm 0.029) \times 10^{-2} $ \\ 
$ -18.225 $ & $ (1.496 \pm 0.022) \times 10^{-2} $ & $ (1.706 \pm 0.024) \times 10^{-2} $ & $ (1.627 \pm 0.023) \times 10^{-2} $ \\ 
$ -18.500 $ & $ (1.358 \pm 0.018) \times 10^{-2} $ & $ (1.519 \pm 0.019) \times 10^{-2} $ & $ (1.465 \pm 0.018) \times 10^{-2} $ \\ 
$ -18.775 $ & $ (1.151 \pm 0.013) \times 10^{-2} $ & $ (1.282 \pm 0.014) \times 10^{-2} $ & $ (1.238 \pm 0.014) \times 10^{-2} $ \\ 
$ -19.050 $ & $ (9.812 \pm 0.102) \times 10^{-3} $ & $ (1.093 \pm 0.011) \times 10^{-2} $ & $ (1.050 \pm 0.011) \times 10^{-2} $ \\ 
$ -19.325 $ & $ (7.996 \pm 0.077) \times 10^{-3} $ & $ (8.617 \pm 0.080) \times 10^{-3} $ & $ (8.459 \pm 0.080) \times 10^{-3} $ \\ 
$ -19.600 $ & $ (6.129 \pm 0.058) \times 10^{-3} $ & $ (6.590 \pm 0.060) \times 10^{-3} $ & $ (6.407 \pm 0.059) \times 10^{-3} $ \\ 
$ -19.875 $ & $ (4.444 \pm 0.043) \times 10^{-3} $ & $ (4.549 \pm 0.044) \times 10^{-3} $ & $ (4.533 \pm 0.043) \times 10^{-3} $ \\ 
$ -20.150 $ & $ (2.938 \pm 0.030) \times 10^{-3} $ & $ (2.849 \pm 0.030) \times 10^{-3} $ & $ (2.887 \pm 0.030) \times 10^{-3} $ \\ 
$ -20.425 $ & $ (1.753 \pm 0.021) \times 10^{-3} $ & $ (1.611 \pm 0.020) \times 10^{-3} $ & $ (1.701 \pm 0.021) \times 10^{-3} $ \\ 
$ -20.700 $ & $ (9.341 \pm 0.133) \times 10^{-4} $ & $ (8.049 \pm 0.129) \times 10^{-4} $ & $ (8.409 \pm 0.129) \times 10^{-4} $ \\ 
$ -20.975 $ & $ (4.358 \pm 0.081) \times 10^{-4} $ & $ (3.380 \pm 0.079) \times 10^{-4} $ & $ (3.834 \pm 0.081) \times 10^{-4} $ \\ 
$ -21.250 $ & $ (1.752 \pm 0.048) \times 10^{-4} $ & $ (1.106 \pm 0.044) \times 10^{-4} $ & $ (1.294 \pm 0.045) \times 10^{-4} $ \\ 
$ -21.525 $ & $ (5.688 \pm 0.269) \times 10^{-5} $ & $ (3.316 \pm 0.248) \times 10^{-5} $ & $ (3.976 \pm 0.251) \times 10^{-5} $ \\ 
$ -21.800 $ & $ (1.418 \pm 0.137) \times 10^{-5} $ & $ (8.716 \pm 1.331) \times 10^{-6} $ & $ (1.094 \pm 0.136) \times 10^{-5} $ \\ 
$ -22.075 $ & $ (4.419 \pm 0.799) \times 10^{-6} $ & $ (2.862 \pm 0.820) \times 10^{-6} $ & $ (3.493 \pm 0.817) \times 10^{-6} $ \\ 
$ -22.350 $ & $ (1.192 \pm 0.448) \times 10^{-6} $ & $ (6.233 \pm 4.366) \times 10^{-7} $ & $ (4.793 \pm 3.358) \times 10^{-7} $ \\ 
$ -22.625 $ & $ (6.726 \pm 3.857) \times 10^{-7} $ & $ (5.336 \pm 5.321) \times 10^{-7} $ & $ (3.564 \pm 3.557) \times 10^{-7} $ \\ 
\hline
\end{tabular}
\end{center}
\label{tab:swml}
\end{table*}

\begin{table*}
\caption{Schechter function fits to the 
2dFGRS galaxy luminosity function for three assumed
cosmological models.
The parameters specify the Schechter functions which, when convolved with
the apparent magnitude measurement errors, give the best 
fits to the SWML estimate of the 2dFGRS galaxy luminosity function. 
The last column lists the integrated luminosity density in solar units 
($M^\odot_{\rm b_J}=5.3$). The contributions to the quoted 
errors on the values of the Schechter function parameters 
have been divided into four distinct categories:
a) The errors directly from STY maximum likelihood estimate
of \Mstar and $\alpha$.
Once combined with the normalization constraint these
induce a corresponding uncertainty in $\Phi^\star$.
b) The contribution due to the uncertainty in the k+e corrections.
c) The uncertainty in the photometric zeropoint.
d) The uncertainty in the normalization due to large scale structure
and residual uncertainty in the incompleteness correction.
If one is interested in the
absolute error in the luminosity function 
these errors should be added in quadrature.
}
\begin{center}
\begin{tabular}{llrrrrrrrrrrrrrrr} 
\multicolumn{1}{l} {$\Omega_0$} &
\multicolumn{1}{l} {$\Lambda_0$} &
\multicolumn{1}{c} {\Mstar$ - 5\logh$} &
\multicolumn{1}{c} {$\alpha$} &
\multicolumn{1}{c} {$\Phi^\star$/$h^3$Mpc$^{-3}$} &
\multicolumn{1}{c} {$\rho_L/h$L$_\odot$ Mpc$^{-3}$} & \\
\hline 
0.3 & 0.7 &  $-19.66\pm0.006^\a\pm0.06^\b\pm0.04^\c$ 
&$-1.21\pm0.01^\a\pm0.02^\b$ 
&$(1.61\pm0.015^\a\pm0.05^\b\pm0.06^\d) \times 10^{-2}$ 
&$(1.82\pm0.17) \times 10^{8}$&\\ 
1   & 0   &  $-19.48\pm0.006^\a\pm0.06^\b\pm0.04^\c$ 
&$-1.18\pm0.01^\a\pm0.02^\b$ 
&$(2.06\pm0.020^\a\pm0.06^\b\pm0.08^\d) \times 10^{-2}$ 
&$(1.92\pm0.19) \times 10^{8}$&\\ 
0.3 & 0   &  $-19.54\pm0.006^\a\pm0.06^\b\pm0.04^\c$  
&$-1.19\pm0.01^\a\pm0.02^\b$ 
&$(1.87\pm0.019^\a\pm0.06^\b\pm0.07^\d) \times 10^{-2}$ 
&$(1.88\pm0.19) \times 10^{8}$&\\ 
\hline			        
\end{tabular} 			 
\end{center}
\label{tab:sty}
\end{table*}

There are two other significant contributions to the uncertainty in
the galaxy luminosity function on an absolute scale.  The first of
these is the zeropoint of the photometry which has an accuracy of
$\pm0.04$~magnitudes. The size of this uncertainty is indicated by the
horizontal error bar plotted in the upper right of each panel of
Fig.\ref{fig:lf_main}, which shows the $\pm3\sigma$ range.  The second
important contribution is the uncertainty in the appropriate
evolutionary correction. Our estimates of the galaxy luminosity
function are at redshift $z=0$ and so rely on an accurate model of the
k+e corrections to transform the measured luminosities, which have a
median redshift of $z_{\rm med}\approx0.1$, to present day values. The
k+e-corrections we use are accurately constrained by the SDSS $\g-\r$
colours, but are nevertheless model dependent at some level. To gauge
the uncertainty in the luminosity function due to this uncertainty we
made SWML LF estimates using k+e-corrections that were increased or
decreased by some factor compared to our standard model. We then
constrained this factor by requiring statistical consistency between
LF estimates made separately for the data above and
below $z=0.1$.  The results of this test for the standard
k+e-correction model were shown in Fig.~\ref{fig:lf6}d, where it can
be seen that the two luminosity functions match accurately.  We find
that if the k+e-corrections are increased or decreased by 18\%, then
the position of the break in the luminosity function between the high
and low redshift samples differs by 1-$\sigma$ (as determined using the
SWML errors). Taking this as an estimate of the uncertainty in the k+e
correction we find that the corresponding uncertainties in the
luminosity function parameters are $\Delta\alpha=0.02$, $\Delta
M^\star=0.06$, and $\Delta\Phi^\star/\Phi^\star = 3\% $.  The
variations in $M^\star$ and \phistar\ are strongly correlated as for a
given value of $M^\star$, \phistar\ is determined using the
normalization constraint provided by the number counts.  This
contribution to the uncertainty in the LF estimates is indicated by
the slanted error bar plotted in the upper right of each panel of
Fig.\ref{fig:lf_main}, which again shows the $\pm 3\sigma$ range.

\subsection{Combined NGP+SGP Estimate}
\label{sec:lf2df2}

The lower panel of Fig.~\ref{fig:lf_main} combines the SGP and NGP
data to give our best estimate of the $\bj$-band galaxy luminosity
function assuming an $\Omega_0=0.3$ and $\Lambda_0=0.7$ cosmology.
The points with $\pm1\sigma$ error bars show the SWML estimate.  Also
shown are two Schechter functions, whose parameter values are
indicated in the legend. The first is a simple STY estimate of the
2dFGRS LF, while the second is obtained by fitting the SWML estimate
by a Schechter function convolved with the distribution of
magnitude measurement errors estimated from Fig.~\ref{fig:mags}. We
see that deconvolving the effect of the magnitude errors causes only a
small reduction in $L^\star$ and $\Phi^\star$. We also see that this
function convolved with the errors (dotted curve) produces a good
match to the SWML estimate. Thus, there is little evidence for the
underlying galaxy luminosity function differing significantly from the
Schechter function form.

The numerical values of these estimates are listed in
Tables~\ref{tab:swml} and~\ref{tab:sty}, along with estimates for
alternative cosmologies.  Note that the SWML estimates refer to the
observed luminosity function, which is distorted by random magnitude
measurement errors.  In contrast, the Schechter function parameters
listed in Table~\ref{tab:sty} refer to the underlying galaxy
luminosity function deconvolved for the effect of magnitude
measurement errors.  In Table~\ref{tab:sty} we have broken down the
errors on the Schechter function parameters into three components. The
first is the statistical error returned by the STY maximum likelihood
method.  The large number of galaxies used in our estimates makes this
statistical error very small and so it is never the dominant
contribution to the overall error. The second error is our estimate of
the error induced by the uncertainty in the
k+e-corrections. This is the dominant contribution to the error in
$\alpha$ and also a significant contributor to the errors in $M^\star$
and \phistar.  The third error given for $M^\star$ in
Table~\ref{tab:sty} is due to the current uncertainty in the 2dFGRS
photometric zeropoint.  This will be reduced when more calibrating CCD
photometry is available. The third error given for \phistar\ is due to
the uncertainty in the galaxy number counts and has contributions from
large-scale structure (3\%) and from the uncertainty in the
incompleteness corrections (2\%).  To determine the overall errors on
an absolute scale these contributions should all be added in
quadrature. For a complete description of the errors one 
also needs to consider the correlations between the different
parameters. For both the contribution to the errors coming from
the uncertainty in the STY parameter estimation and from the uncertainty in the
k+e-correction a steeper faint end slope, $\alpha$, correlates
with brighter $M^\star$. This, in turn, is correlated with \phistar\
as the number count constraint implies that a brighter $M^\star$
will produce a lower \phistar.
In each case the  correlation coefficient is large, $R\approx0.8$.
The uncertainty in the photometric zeropoint affects only 
$M^\star$, while the uncertainty in the number count constraint 
affects only \phistar. This reduces the correlation between the parameter 
estimates.
The final column in Table~\ref{tab:sty} lists the implied
$z=0$ luminosity density in solar units. The error quoted on this
quantity was computed by propagating all the previously mentioned
errors. An alternative estimate of the error can be obtained by
by estimating the luminosity density independently
from the NGP and SGP data. This gives 
$\rho_L=2.04\times10^8 h$L$_\odot$Mpc$^{-3}$ (NGP) 
and~$1.64\times10^8 h$L$_\odot$Mpc$^{-3}$ (SGP) indicating a
very similar mean luminosity density and uncertainty.

 The Schechter function parameters listed in Table~\ref{tab:sty} for
the $\Omega_0=0.3$, $\Lambda_0=0.7$ cosmology differ slightly from
those in Madgwick \etal (\shortcite{madgwick01}).  This is to be
expected as the Madgwick \etal luminosity functions are not corrected
for evolution. That paper focused on the dependence of the luminosity
function on spectral type.  Adopting the average k-correction of
Madgwick \etal and using this in place of our k+e-correction on our
larger sample (the Madgwick \etal sample is truncated at $z=0.15$), we find
luminosity function parameters very close to those of Madgwick \etal
(\shortcite{madgwick01}).  The remaining, very small differences are
accounted for by slightly differing models for the magnitude errors
and the adopted normalizations.

\section{Comparison with Independent Luminosity Function Estimates}
\label{sec:lf.compare}

\begin{figure*}
\epsfxsize=15.5 truecm \epsfbox[20 40 530 745]{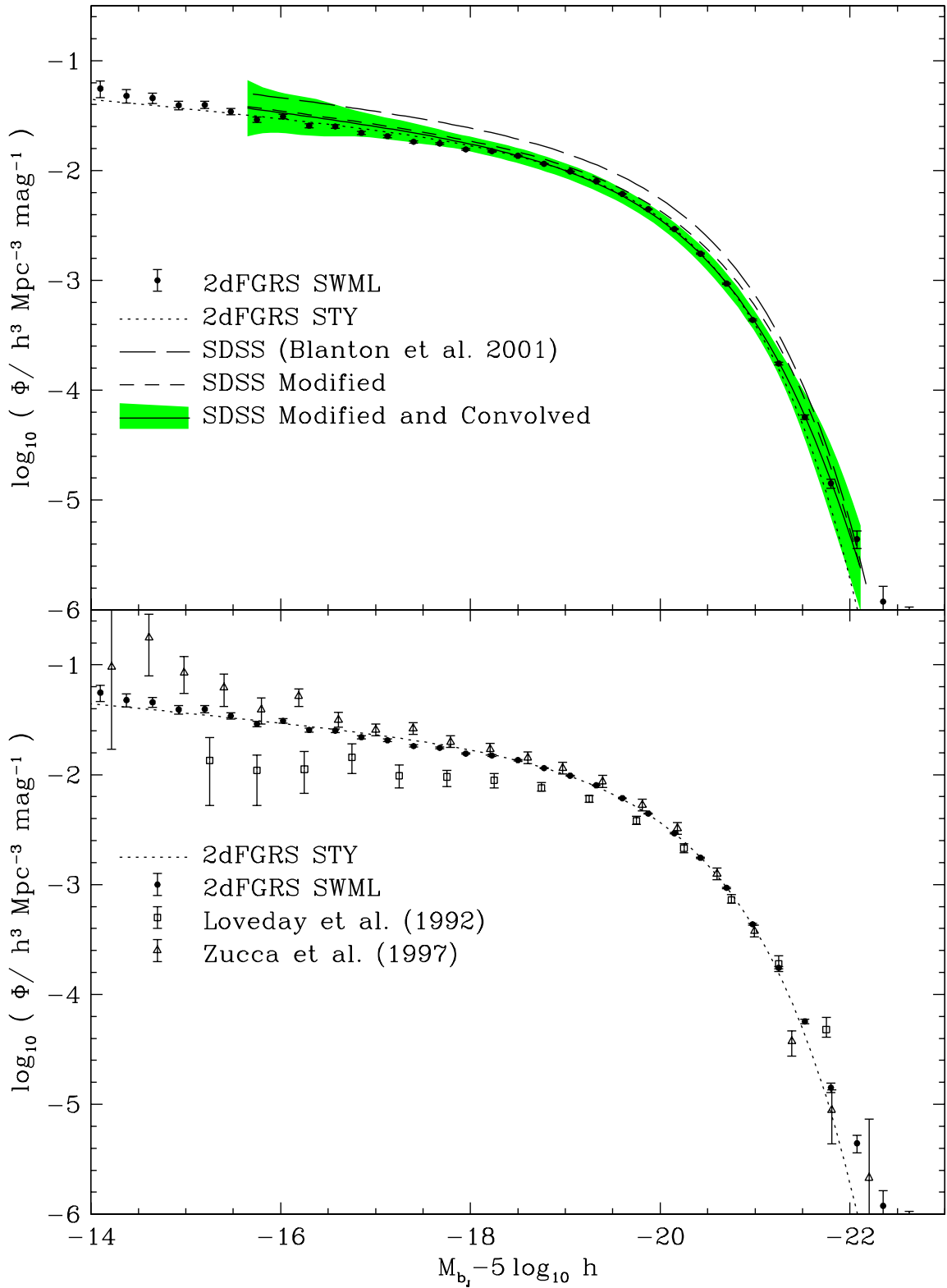}
\caption{Comparison of the 2dFGRS $\bj$-band luminosity function
with estimates from the SDSS and the earlier estimates of
Loveday \etal (1992) and Zucca \etal (1997).
}
\label{fig:lf_two}
\end{figure*}

In Fig.~\ref{fig:lf_two} we compare the STY and SWML estimates of the
$\bj$-band LF from the combined NGP+SGP 2dFGRS sample
defined by $17<\bj<19.2$ and $z<0.25$ (shown in Fig~\ref{fig:lf_main})
with estimates from other surveys. The upper panel compares 2dFGRS
with various estimates made from the SDSS.  In this comparison we
again assume an $\Omega_0=0.3$, $\Lambda_0=0.7$ cosmology. Blanton
\etal (\shortcite{blanton}) presented an estimate of the $\bj$-band LF
for the case of $\Omega_0=1.0$. We do not use this, but instead
estimate the $\bj$-band LF for our adopted cosmology using the
$\g$-band LF computed by Blanton \etal (\shortcite{blanton}) for the
$\Omega_0=0.3$, $\Lambda_0=0.7$ cosmology and the typical $B-V$ galaxy
colour.  Using the colour equations of Fukugita \etal
(\shortcite{fukugita96}), and assuming $\bj= B-\beta(B-V)$, one finds
$\bj= \g+0.12+(0.44-\beta)(B-V)$. Blanton \etal \shortcite{blanton} 
assumed $\beta=0.35$, based on the work of Metcalfe \etal 
(\shortcite{metcalfe}), and contrary
to the commonly-used value of $\beta=0.28$ (Blair \& Gilmore \shortcite{bj}).
Thus, an estimate of the $\bj$-band LF can be made by simply taking
the $\g$-band estimate and shifting the magnitudes using this equation
with $B-V=0.94$, this being the mean colour measured for galaxies
brighter than $\bj=19$ in the SDSS sample.  This procedure can been
seen to work quite accurately: when applied to the $\Omega_0=1$
$\g$-band LF parameters given in table~2 of Blanton \etal
(\shortcite{blanton}), it reproduces the corresponding $\bj$
parameters given in their Fig.~23.  Taking $\beta=0.35$ and applying
this procedure for the $\Omega_0=0.3$, $\Lambda_0=0.7$ cosmology gives
$M_\bj^\star-5\logh=-19.82$, $\alpha=-1.26$ and $\Phi^\star=2.06
\times 10^{-2} h^3$Mpc$^{-3}$ . This Schechter function is shown by
the long dashed curve in the upper panel of Fig.~\ref{fig:lf_two}.  As
discussed by Blanton \etal (\shortcite{blanton}), this estimate is
incompatible with the 2dFGRS estimate and predicts a significantly
higher luminosity density than we find.

The short dashed line in the upper panel of Fig.~\ref{fig:lf_two}, a
Schechter function with $M_\bj^\star-5\logh=-19.68$, $\alpha=-1.26$
and $\Phi^\star=1.56 \times 10^{-2} h^3$Mpc$^{-3}$, is the result of
making three modifications to the Blanton \etal (\shortcite{blanton})
curve.  First, we have shifted $M_\bj^\star$ by $0.066$ magnitudes as
is appropriate if one adopts the Blair \& Gilmore (\shortcite{bj})
colour equation $\bj= B-0.28(B-V)$ rather than $\bj= B-0.35(B-V)$ used
by Blanton \etal (\shortcite{blanton}). The latter is actually ruled
out by the empirical relations found by matching the 2dFGRS catalogue
with either the EIS or SDSS which are instead consistent with the
former.  Second, we have shifted $M_\bj^\star$ by a further $0.058$ to
take account of the zeropoint offset between the SDSS and 2dFGRS
photometry that we found in Section~\ref{sec:photometry}
(Fig.~\ref{fig:mags}c). Finally, we have reduced $\Phi^\star$ by
$24\%$, the reduction required for this luminosity function to
reproduce the mean 2dFGRS number counts at $\bj=19.2$ assuming our
standard k+e-correction model.  We note that Yasuda \etal
(\shortcite{yasuda}) also found a value of \phistar\ significantly
lower than Blanton \etal when they normalized the SDSS $\g$-band
luminosity function using the SDSS galaxy counts.  The Yasuda \etal
estimate is still higher than our value for two reasons.
First the Yasuda \etal counts are 8.4\% high as they include extended 
sources that do not satisfy all the galaxy selection criteria used
in the main SDSS galaxy sample. Second although the SDSS galaxy
counts agree with 2dFGRS in the area of overlap, this smaller area
($173\,$deg$^2$) has a 5\% higher density of galaxies than the full
area ($1834\,$deg$^2$) covered by the 2dFGRS survey (see
Section~\ref{sec:lf.norm}).  This modified SDSS Schechter function is
in near perfect agreement with the Schechter function estimated from
the 2dFGRS.

At the brightest magnitudes, the 2dFGRS SWML estimate is above both
the 2dFGRS STY estimate and the modified SDSS Schechter function
estimate.  As we have seen, the main reason for this is that magnitude
measurement errors in the 2dFGRS have a significant effect on the
bright end of the luminosity function, but little effect around
$M^\star$ and fainter.  The solid curve surrounded by the shaded
region shows the result of convolving the modified SDSS estimate with
the model of the 2dFGRS magnitude errors shown in Fig.~\ref{fig:mags}.
The shaded region indicates the statistical error on the SDSS estimate
and was read from figure~6 of Blanton \etal (\shortcite{blanton}).
Comparing this with the 2dFGRS SWML estimate we see that the two are
perfectly consistent, with the larger 2dFGRS sample having
significantly smaller statistical errors.

We have seen that after taking into account the zeropoint photometric
offset and the error in the colour equation, the only significant
difference between the LF estimates of Blanton \etal
(\shortcite{blanton}) and the 2dFGRS is a difference in \phistar. This
difference arises not because the density of galaxies is higher in
SDSS than 2dFGRS (the counts agree to 5\%), but because of the
different methods used to constrain \phistar.  Blanton \etal used the
method of Davis \& Huchra (\shortcite{davishuchra}) which weights
galaxies as a function of redshift in order to obtain a minimum
variance estimate of the galaxy density. This method gives more weight
to galaxies at high redshift than the method based on normalizing to
the counts. It results in a smaller statistical error in the
normalization, but at the same time renders the result more dependent
on the accuracy of the evolutionary correction.  We have seen in
Section~\ref{sec:lf2df1} that, even with the low redshift constraint
provided by the galaxy counts, the uncertainty in \phistar\ due to the
uncertainty in the k+e correction is significant. With the Davis \&
Huchra weighting this uncertainty becomes dominant. The analysis by
Blanton \etal did not take account of evolution -- only k-corrections
were applied -- and this appears to have given rise to an artificially
high estimate of \phistar\ in the $\g$-band.  We conclude that, when
normalized in the same way, there is excellent agreement between the
SDSS and 2dFGRS luminosity functions and that the dominant remaining
uncertainty in the present day $\bj$-band LF is due
to residual uncertainties in evolutionary corrections.

\begin{figure*}
\hbox{ \epsfxsize=8.5 truecm \epsfbox[10 75 540 750]{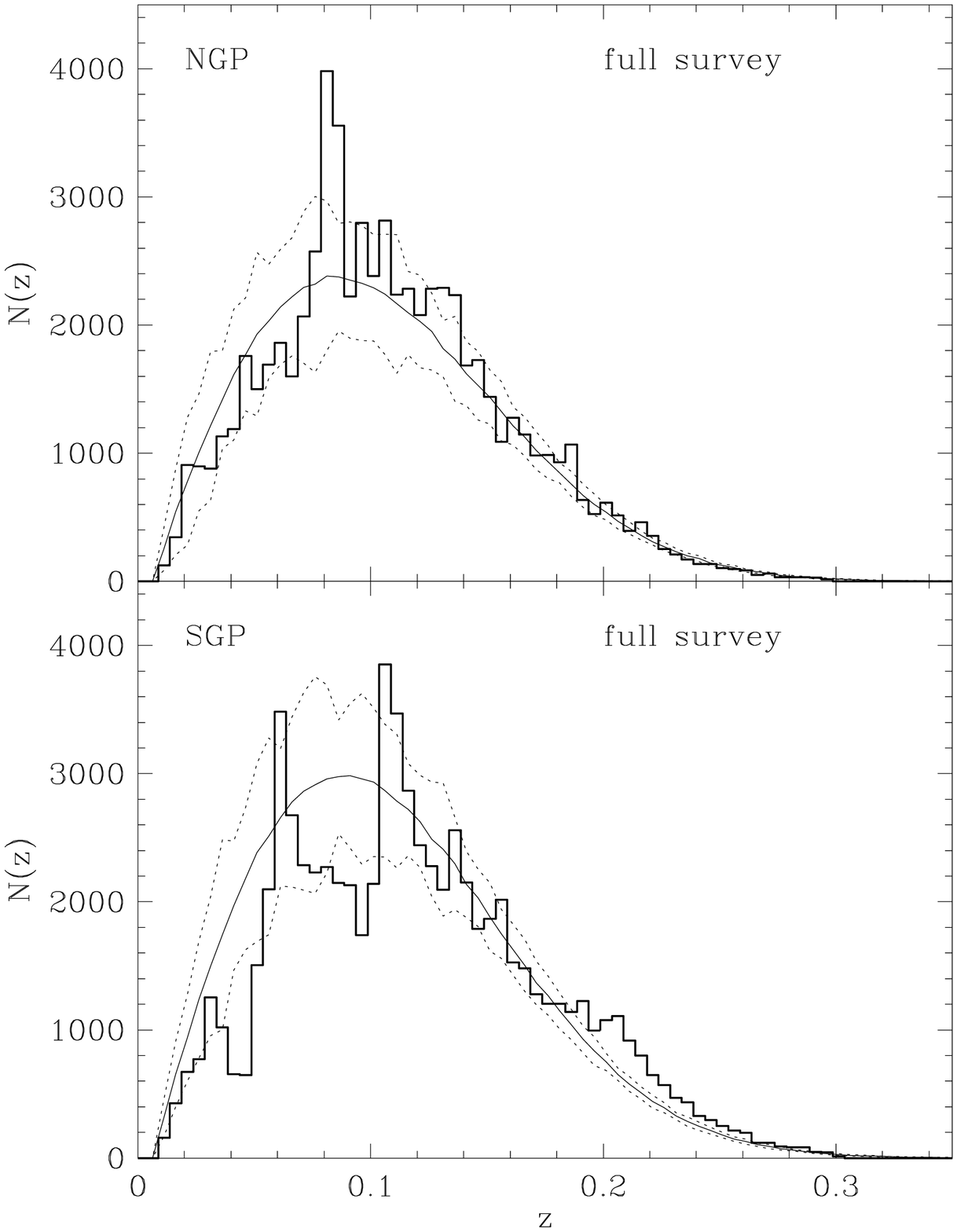}
       \epsfxsize=8.5 truecm \epsfbox[10 75 540 750]{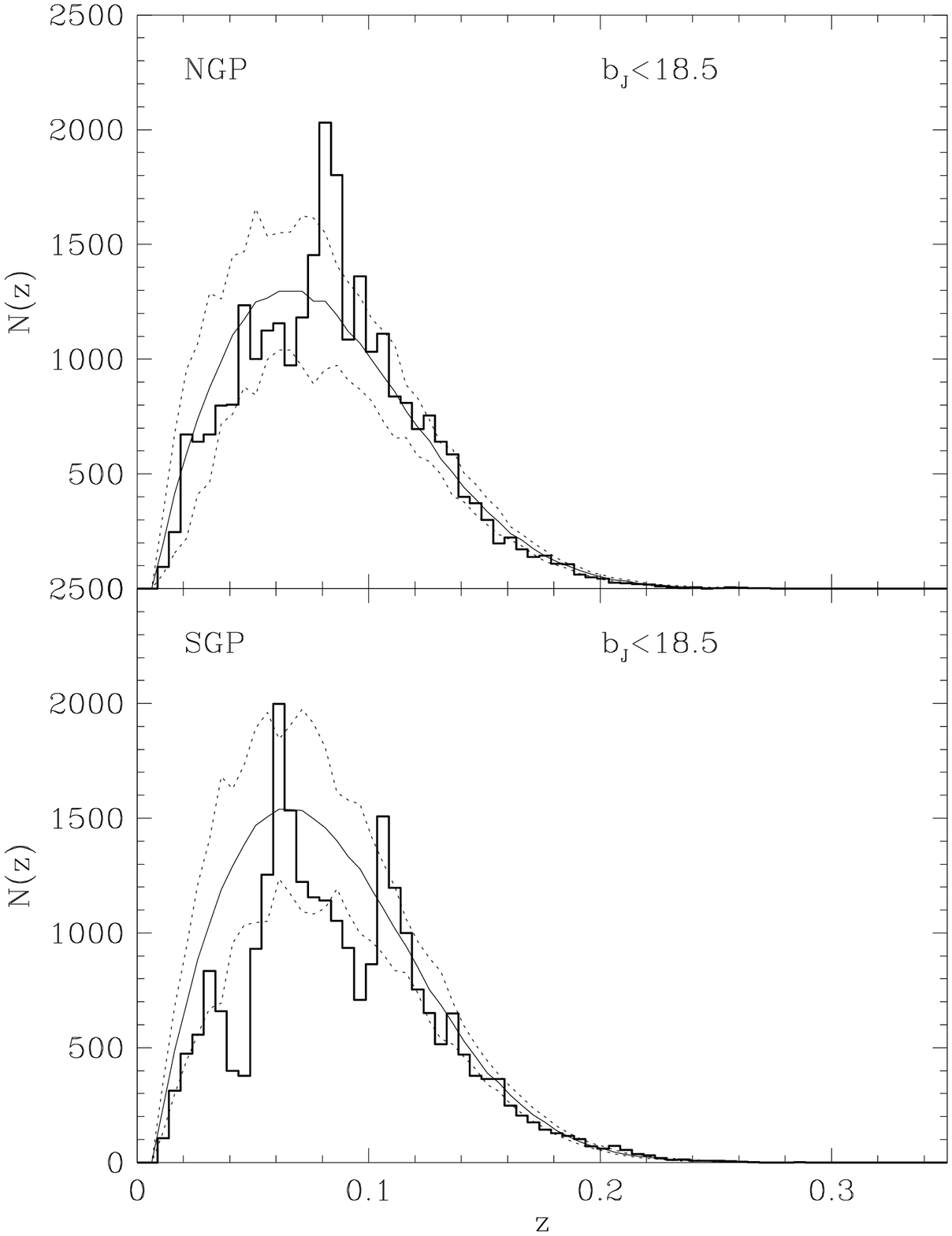}
}
\caption{Redshift distributions in the 2dFGRS and mock catalogues.
The histograms show the observed redshift 
distribution in the NGP and SGP regions of the 2dFGRS. The left-hand panels 
are to the full depth of the survey while the 
right-hand panels include only galaxies brighter than $\bj=18.5$.
The smooth solid curves show
the predicted redshift distributions based on our Schechter function
estimate of the galaxy luminosity function,
including the magnitude measurement errors, the variation in
the survey magnitude limit and the dependence of completeness on
apparent magnitude. The dotted lines indicate the rms variation in
the redshift histograms within our ensemble of 22 mock galaxy catalogues.
}
\label{fig:dndz}
\end{figure*}

The lower panel of Fig.~\ref{fig:lf_two} compares the 2dFGRS result
with the earlier estimates of Loveday \etal (\shortcite{loveday}) and
Zucca \etal (\shortcite{zucca}). We see that the Zucca \etal estimate
agrees well with 2dFGRS although it has statistical errors that are
much larger.  The main difference with the luminosity function of
Loveday \etal (\shortcite{loveday}) is its lower \phistar.  Both the
2dFGRS and Loveday \etal
estimates are based on catalogues extracted from the APM survey.
However, the Loveday \etal sample is much brighter and almost disjoint
from the sample analyzed in this paper.  As we have seen, the bright
galaxy number counts in the SGP drop below model predictions
extrapolated from fainter magnitudes (Maddox \etal
\shortcite{apm_counts} and Section~\ref{sec:lf.norm}) and it is
therefore not surprising that Loveday found a lower value of \phistar.
Similarly, the flatter faint end slope that they find might be attributed,
at least in part, to small volume effects. This explanation has been
argued by Zucca \etal, who find they are able to accurately reproduce  
the Loveday \etal result if they analyze only the subset of their
galaxy sample brighter than the $\bj<17.15$ limit of Loveday \etal
(\shortcite{loveday}).

\section{The 2\lowercase{d}FGRS Selection Function}
\label{sec:selfun}

The luminosity function we have derived, combined with the maps
defining the survey magnitude limit (see figure~13 \cite{colless01}),
redshift completeness (see Fig.~\ref{fig:mask}) and $\mu$-parameter
(see Fig.~\ref{fig:mumask}) specify the complete 
selection function of the 2dFGRS.\footnote{
       The only significant aspects of the 2dFGRS selection function
       ignored in this description are surface brightness issues
       (see Cross \etal \shortcite{cross01}) and the undersampling of close
       galaxy pairs induced by the mechanical limits on the positioning
       of the optical fibres that feed the 2dF spectrographs.
       Note that as the 2dF fields overlap, not all close
       galaxy pairs are missed. We have found that when making
       estimates of galaxy clustering an accurate
       way of dealing with this incompleteness is to assign the
       weight of the missed galaxies to neighbouring galaxies
       with redshifts. We typically distribute the weight of a
       missed galaxy 
       between its 10 nearest neighbours and find that this produces
       accurate clustering estimates on scales greater than
       1.5~arcmin or $\sim0.2h^{-1}$Mpc (\cite{norbergC}).
} 
It is interesting to compare the redshift distribution implied
by this selection function with the measured distribution. 
Note that the luminosity function estimators we employed are insensitive
to clustering and so the information contained in the redshift
distribution of the galaxies has not been used in determining
our model of the selection function.

\begin{figure*}
\hbox{ \epsfxsize=8.5 truecm \epsfbox[10 75 540 750]{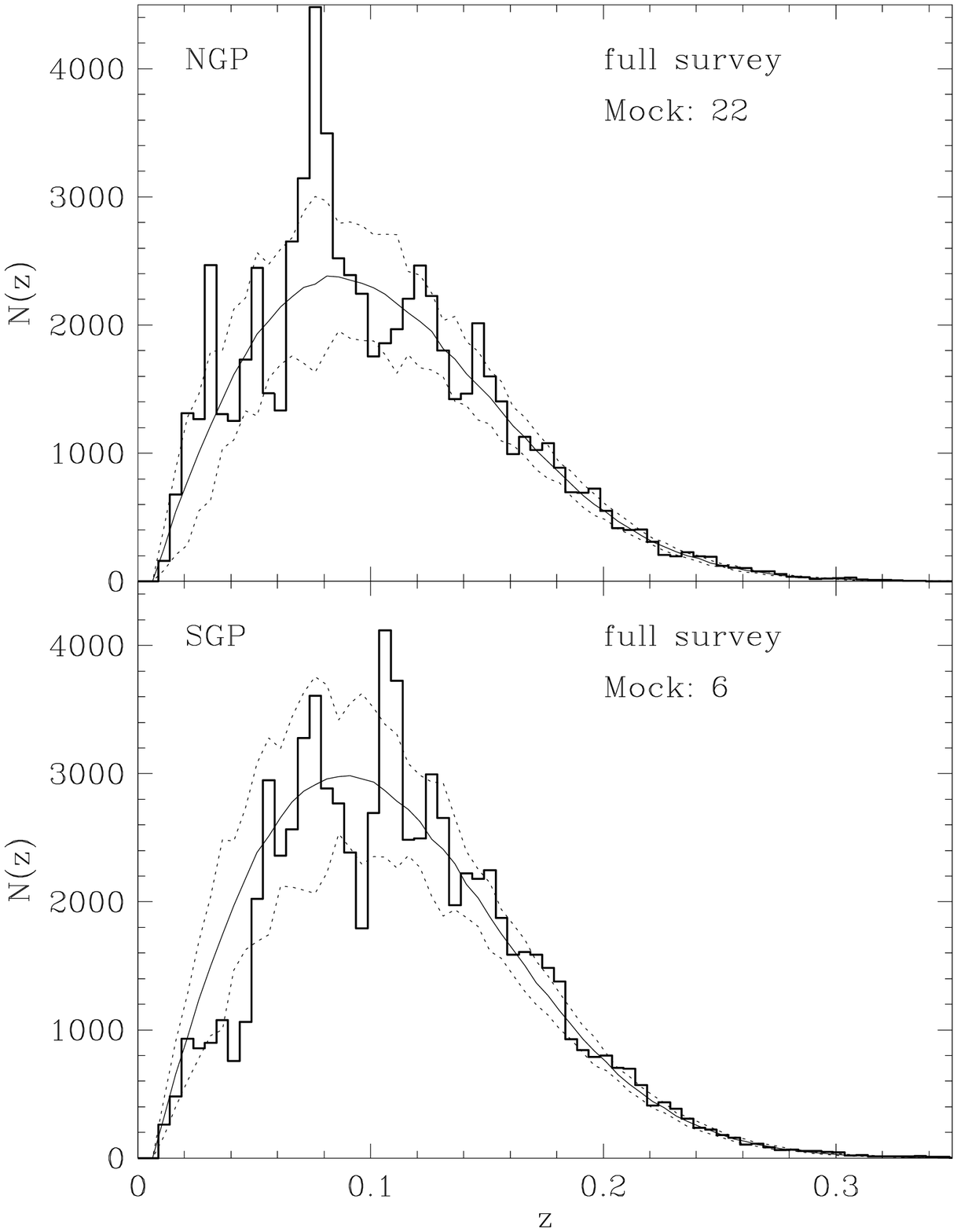}
       \epsfxsize=8.5 truecm \epsfbox[10 75 540 750]{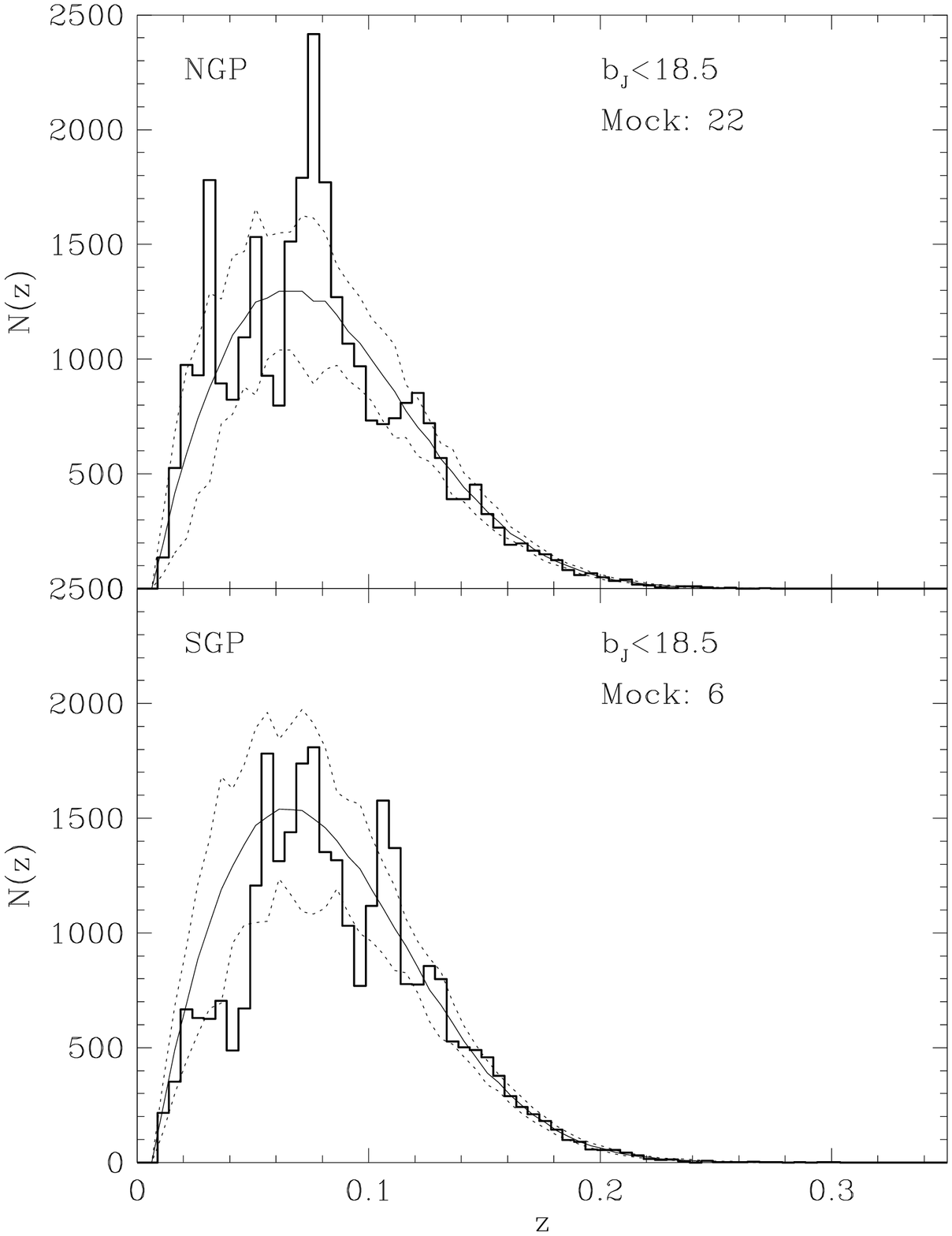}
}
\caption{As Fig.~\ref{fig:dndz}, but for two selected mock catalogues
rather than the genuine 2dFGRS.
}
\label{fig:mockdndz}
\end{figure*}

In Fig.~\ref{fig:dndz} we compare the smooth redshift distribution
predicted by our model of the 2dFGRS selection function with the
observed distribution. The left-hand panels show the redshift
distributions for the full 2dFGRS survey split into the SGP and NGP
regions. The right-hand panels show the distributions only for
galaxies brighter than $\bj=18.5$.  The dotted lines indicate the rms
variation in the redshift histograms found in our 22 mock 2dFGRS
catalogues.  Gravitational clustering produces a pattern of galaxy
clustering that is non-gaussian, composed of voids, walls, filaments
and clusters (\eg see figures~8 to~15 of Cole \etal
\shortcite{oldmocks} for mock 2dFGRS and SDSS cone plots). As a
result, the rms variation in the $N(z)$ distribution does not give an
adequate description of the variation seen in the mock catalogues. For
this reason we show in Fig.~\ref{fig:mockdndz} two examples of the
redshift distributions found in our ensemble of mock catalogues.  From
these we see that the few large spikes present in the $N(z)$ of the
2dFGRS data are common features in the mock catalogue redshift
distributions.

The redshift distribution in the 2dFGRS NGP has a large spike close to
the peak of the selection function and otherwise lies within
1-$\sigma$ of our smooth selection function. Thus, the density field
in the NGP strip looks in no way unusual when compared to the
expectation in the standard $\Lambda$CDM ($\Omega_0=0.3$,
$\Lambda_0=0.7$) universe.  In contrast, the density field in the SGP
appears more extreme.  Focusing first on the redshift distribution
below $z<0.2$, we see that the observed galaxy density is nearly
always below the mean density predicted by the selection
function. This behaviour is consistent with the steep APM galaxy
number counts, first noted by Maddox \etal (\shortcite{apm_counts}),
and discussed in Section~\ref{sec:lf.norm} above. A lower than average
galaxy density over such a large range of redshift is certainly not a
common occurrence.  However, as illustrated by the example of the
mock SGP plotted in Fig.~\ref{fig:mockdndz}, which in many respects is
quite similar to the observed 2dFGRS SGP, comparable variations do
occur in the mock $\Lambda$CDM catalogues.  The two examples plotted
in Fig.~\ref{fig:mockdndz} were not chosen at random, but as we only
have 22 mocks to choose from, they do not represent extreme
possibilities.

The 2dFGRS SGP strip also appears to show an overdensity, relative  to
the mean implied by the selection function, in the redshift range
$0.2<z<0.25$.  As the volume contributing to this redshift interval is
very large, a variation as extreme as this is very unlikely. It
therefore seems implausible that this perturbation in $N(z)$ is due 
solely to large-scale structure. There are some structures at this
redshift that contribute to the excess, but even if they are excised
the $N(z)$ remains higher than the model.
At $z>0.2$ the only galaxies which make it into the 2dFGRS
are one to two magnitudes brighter than $M^\star$, where the galaxy
luminosity function is very steep. Thus, a small shift in magnitude 
can result in a large change in the number of galaxies brighter than 
the survey magnitude limit. We cannot reject the possibility of a small
offset between the absolute calibration of the NGP and SGP; indeed,
Section~\ref{sec:sdss} 
has indicated an offset of $0.058$~magnitudes -- assuming both EIS and SDSS
photometry to be perfect. This is in the sense required to understand
Fig.~\ref{fig:dndz}, i.e. the SGP is effectively deeper than the NGP. Any true
offset cannot be much larger than this, otherwise it would spoil the
good agreement of the STY LFs in Fig 11. Nevertheless,
an NGP--SGP offset of between 0.05 and 0.1~magnitudes would yield a better
match to $N(z)$ at $z>0.2$, and this possibility must be borne
in mind, pending further tests against CCD data.
Another possibility that needs further investigation is that the random
magnitude measurement errors become larger for faint objects at high
$z$. A trend of this sort is not evident in the comparison we have
made between 2dFGRS and SDSS EDR magnitudes in Fig~\ref{fig:mags}, but
this comparison pertains to the NGP only. For now one should be careful, as
we have been in previous papers, to ensure that large-scale clustering
results are not strongly influenced by this feature. For instance,
the estimate of the large scale galaxy power spectrum 
in  Percival \etal (\shortcite{percival}) used separate selection
functions, which empirically matched the high-$z$
$N(z)$ in both NGP and SGP.

\section{Discussion and Conclusions}
\label{sec:concl}

We have described the calibration of the 2dFGRS photometry and
used the CCD data of the SDSS EDR (\cite{edr}) to assess the
accuracy and completeness of the 2dFGRS photometric catalogue, which
is based on APM scans of the UKST photographic plates
(\cite{apmII}). We find that the measurement errors in the APM
magnitudes are in agreement with previous estimates, having a
1-$\sigma$ spread (robustly estimated) of $0.164$~magnitudes. We find
a small zeropoint offset between the SDSS EDR and the 2dFGRS
photometry of $\vert\Delta\vert=0.058$ and no evidence for any scale
error in the magnitude calibration in the range $17<\bj<19.5$. As more
calibrating data become available, the accuracy of both the 2dFGRS and
SDSS photometric zeropoints should be improved.  We find that compared
to the SDSS photometric catalogue, the 2dFGRS parent catalogue is
$91\pm2$\% complete.  This agrees with the original estimates based on
the accuracy of star-galaxy classification in the APM catalogue
(\cite{apmI}). The reasons behind the $9\pm2$\% of galaxies that are
missed are investigated in more detail in Cross \etal
(\shortcite{cross01}), who compare the 2dFGRS parent catalogue with
the MGC, a deep, wide area B-band CCD imaging survey
(\cite{mgc}). They find that mis-classification (\eg galaxies
incorrectly classified as merged images or stars) is the largest cause
of incompleteness, but also a small population of low surface
brightness galaxies is missed.

Making simple statistical corrections for incompleteness, magnitude
measurement errors and uncertainties in modelling evolution and
k-corrections, we find that the true $z=0$ galaxy luminosity function
is accurately described by a Schechter function with parameters:
$M^\star_\bj-5\logh=-19.66\pm0.07$, $\alpha=-1.21\pm0.03$ and
\phistar$=(1.61\pm0.08)\times 10^{-2} h^3$Mpc$^{-3}$ (assuming an
$\Omega_0=0.3$, $\Lambda_0=0.7$ cosmology). With over 110\,500
redshifts, the statistical errors in our estimate are negligible
compared to the systematic errors (\ie uncertainties that cause an
overall shift of the luminosity function) from fluctuations produced
by large-scale structure and by the uncertainty in the evolutionary
corrections. Our quoted errors include estimates of these
uncertainties, the former derived from extensive, realistic mock
catalogues.

Taking account of the photometric zeropoint difference, random
magnitude measurement errors, and using an accurate colour equation, we
find very good agreement between the form of the $\bj$-band LF
inferred from the SDSS data and the 2dFGRS estimate.  Also, in the
area of overlap, the 2dFGRS and SDSS galaxy counts agree at
$\bj=19.2$. This is the magnitude at which we use the counts to
normalize our luminosity function.  Thus, when normalized in the same
way, 2dFGRS and SDSS $\bj$-band LF estimates agree with great
accuracy.  Blanton \etal (\shortcite{blanton}) 
reached a different
conclusion principally because they used an inaccurate colour equation
to convert from SDSS wavebands to $\bj$ and did not take account of
galaxy evolution.

The integrated $z=0$ $\bj$-band luminosity density implied by the
2dFGRS LF is $(1.82\pm0.17) \times 10^{8} h$ L$_\odot$Mpc$^{-3}$. This
is in good agreement with earlier estimates from the 2dFGRS presented
in Folkes \etal (\shortcite{folkes}) and Madgwick \etal
(\shortcite{madgwick01}) although neither of these estimates took
account of the small effect of modelling evolution and the Folkes
\etal estimate assumed an $\Omega_0=1$ cosmology.  The dependence of
the estimated luminosity density on corrections made for low surface
brightness galaxies is studied in Cross \etal (\shortcite{cross01}),
which supercedes Cross \etal (\shortcite{cross00}). The earlier paper
modelled the 2dFGRS magnitudes as gaussian-corrected isophotal
magnitudes, in the same way as Blanton \etal
(\shortcite{blanton}). Ignoring the manner in which the 2dFGRS
magnitudes are calibrated using deeper CCD magnitudes led to an
overestimate of the amount of light lost.  Cross \etal
(\shortcite{cross01}) show that the luminosity function estimated here
agrees well with an estimate from the deeper MGC catalogue and that
dependencies on surface brightness only significantly affect the
luminosity function fainter than $\Mag=-16.5$. Consequently
the systematic effect on the estimated luminosity density is only 5 to~10\%.

Wright (\shortcite{wright}) has  highlighted that the luminosity
density measured in the optical bands by the SDSS (\cite{blanton}),
combined with a simple model for the expected mean spectrum, predicts a
luminosity density in the $\K$-band a factor of $2.3$ greater than the
value measured in the joint analysis of 2MASS (\cite{2mass}) and
2dFGRS presented in Cole \etal (\shortcite{irlf}). Note that
Kochanek \etal (\shortcite{kochanek}) found a very similar
$\K$-band luminosity density to that found by
Cole \etal (\shortcite{irlf}), but their estimate used 2MASS
isophotal magnitudes for which the correction to total magnitudes
is more uncertain. Even if the SDSS
luminosity densities were to be revised downwards to agree with the
2dFGRS in the $\bj$-band, the discrepancy in the $\K$-band would only
be reduced to a factor of $1.6$. Furthermore, the correction for
longer wavelength bands is likely to be smaller than that we have
inferred for the $\g$-band.  Thus, a puzzling factor of approximately
$1.8$ to $2$ remains between the $\K$-band luminosity density measured
from 2MASS and that inferred by extrapolation from the optical bands.

Wright (\shortcite{wright}) speculated that the 2MASS fluxes could
be grossly underestimated. This possibility is ruled out by the
comparison of 2MASS magnitudes with those from deeper, high resolution
$\K$-band images. Cole \etal (\shortcite{irlf}) compared 2MASS
magnitudes with Kron magnitudes measured for the same objects by 
Loveday (\shortcite{lovedayk}) and found that only a $0.06$~magnitude
correction was needed to bring them into agreement. A somewhat larger
larger offset, but still only 20\%, has been argued for by
Andreon (\shortcite{andreon}). Furthermore, combining the 2MASS
and SDSS EDR magnitudes for matched objects, we find optical to near
infrared colours which, on average, agree well with the mean galaxy
spectrum adopted by Wright. A second speculation made by Wright was
that perhaps the 2MASS extended source catalogue is incomplete and
misses a significant fraction of the galaxies that SDSS detects.  This
is also appears unlikely. The assessment of the completeness of 2dFGRS
compared to 2MASS presented in Cole \etal (\shortcite{irlf}), together
with the assessment of the 2dFGRS completeness with respect to the
SDSS presented here, shows that the 2MASS and SDSS source densities
agree to about 2\%.

The most likely cause of the discrepancy between the $\K$-band and
extrapolated optical luminosity densities is large-scale structure.  
Since the 2MASS
survey has a much brighter limiting magnitude than either the 2dFGRS or
SDSS, their luminosity functions are not normalized within the same
volume. Cole \etal (\shortcite{irlf}) normalized their $\K$-band LF
using an estimate of the counts from a small, $184\,$deg$^2$ area (Jarrett
\etal in preparation) and an indirect estimate from the approximately
$619\,$deg$^2$ of overlap between 2MASS and 2dFGRS. The second estimate
is perhaps not highly accurate, because it requires
an estimate of the effective area of sky in the intersection of
the 2dFGRS and 2MASS. This is not trivial to obtain because a map of
the 2MASS sky coverage is not yet available. Cole \etal
(\shortcite{irlf}) estimated that large-scale structure would
cause a 15\% variation in the number counts within a $619\,$deg$^2$
area. Our mock catalogues, modified to mimic the selection criteria of the
2MASS, show that the rms variation in the counts over a
$184\,$deg$^2$ area is, significantly larger, 19\%. Thus, it will be very
interesting to derive the $\K$-band counts over a larger area, which
should soon become possible with a more complete 2MASS catalogue, to see
whether the estimates of the J and $\K$-band luminosity densities and the
inferred stellar density need to be revised.

We have described maps that define the redshift completeness of the
current 2dFGRS catalogue and the weak dependence of the degree of completeness
on apparent magnitude. These, together with the luminosity function and
a map of the survey magnitude limit, provide a complete 
description of the 2dFGRS selection function. We have created mock galaxy
catalogues from  cosmological N-body simulations using this
description of the selection function. Comparison of these with the
observed data indicates that, in general, the data are well described
by our selection function and exhibit fluctuations that are typical
of those expected in the standard $\Lambda$CDM cosmology.

\section*{Acknowledgements}
The data used here were obtained with the 2 degree field facility on the 
3.9m Anglo-Australian Telescope (AAT). We thank all those involved in 
the smooth running and continued success of the 2dF and the AAT. 
We thank Valerie de Lapparent
for kindly making available the ESO-Sculptor photometry.
We thank the anonymous referee for many useful comments.

\appendix

\section{Redshift Incompleteness in the 2\lowercase{d}FGRS}
\label{app:maps}

\begin{figure*}
\epsfxsize=18.5 truecm \epsfbox{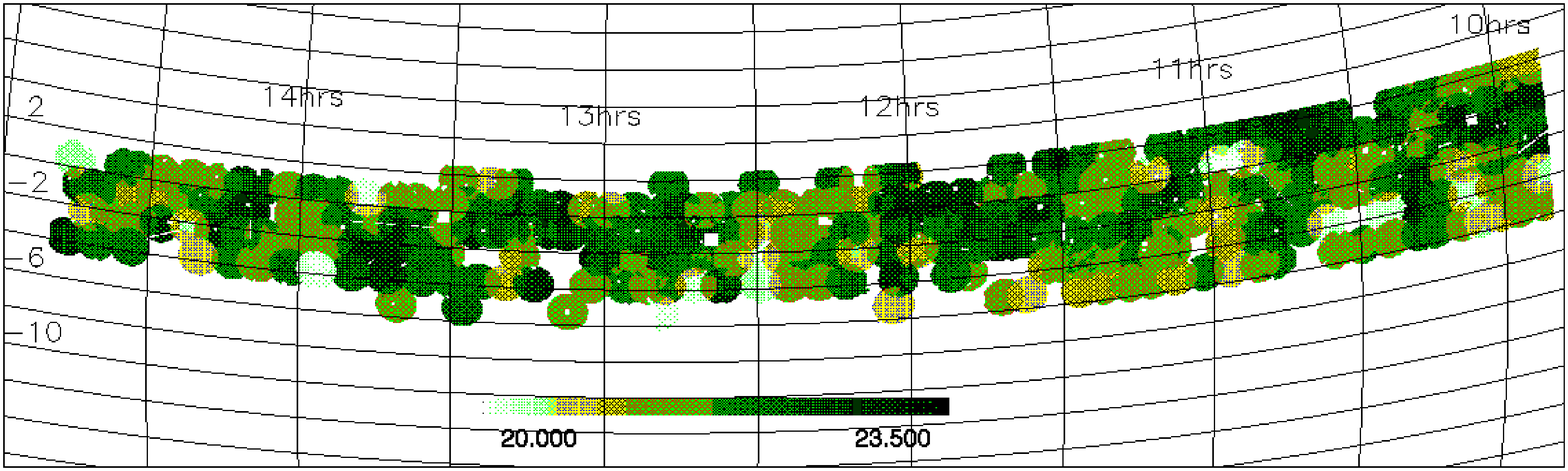}
\epsfxsize=18.5 truecm \epsfbox{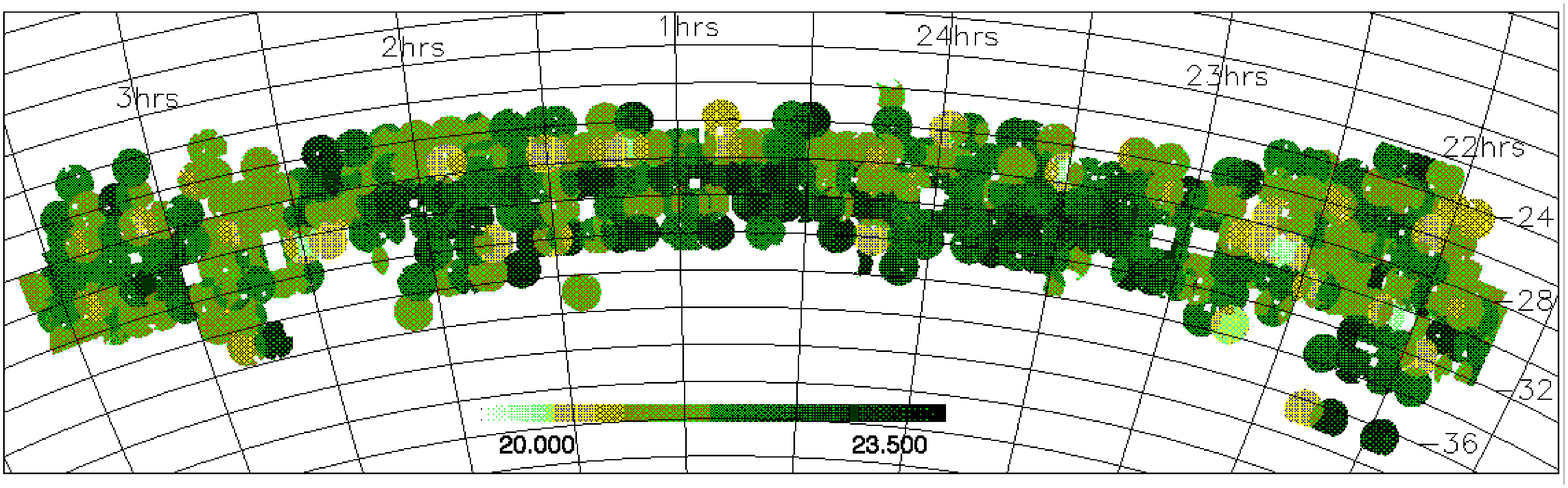}
\caption{Map showing the variation of the parameter $\mu$
with position on the sky. The dependence of the redshift completeness
on apparent magnitude is accurately described by the fitting 
function $c_z(\bj,\mu)= 0.99 \left(1-\exp[\bj-\mu] \right)$.
}
\label{fig:mumask}
\end{figure*}

When complete, the fraction of redshifts measured should be uniformly
high across the full area of the 2dFGRS. However, at this intermediate
stage, when only a subset of the target 2dF fields have been
observed, the fraction of redshifts measured varies considerably with
position. As detailed in Section~8 of Colless \etal
(\shortcite{colless01}), this variation is best quantified by dividing
the survey into sectors (labelled by an angular position $\btheta$)
defined by the overlaps of the target 2dF fields. Within each of these
sectors one can calculate the fraction $R(\btheta)$ of the parent
catalogue galaxies whose redshifts have been measured. It is this
completeness map, pixellated for convenience, that is shown in
Fig.~\ref{fig:mask}.

In contrast to most previous redshift surveys, the 2dFGRS is so large
that residual small systematic errors can begin to dominate over
statistical errors. For this reason, we have developed a quantitative
description of the dependence of the completeness on apparent
magnitude. Note that 76\% of the observed fields have an overall
completeness of greater than 90\% (this should increase with time as
some of the lower completeness fields are re-observed) and so
generally incompleteness and its dependence on apparent magnitude are
small. In Section~8.3 of Colless \etal (\shortcite{colless01}), we
showed that for each observed field the dependence of the redshift
completeness on apparent magnitude could be described by a one
parameter function (see figure~16 of \cite{colless01})
\begin{equation}
c_z(\bj,\mu_i)= 0.99\, \left( 1-\exp[\bj-\mu_i] \right) .
\end{equation}
Here $\bj$ is the apparent magnitude and
$\mu_i$ is the value of the  parameter for field $i$.
In each sector, the targeted galaxies are
split between several fields and so one must define an appropriately
averaged value, $\mu(\btheta)$, for each sector.  This can be
derived by writing the magnitude-dependent redshift incompleteness 
of a sector $c_z(\bj,\mu(\btheta))$ as a weighted sum of the
completeness of its $N_{\rm F}(\btheta)$ component fields,
\begin{equation}
c_z(\bj,\mu[\btheta]) = \sum_{i=1}^{N_{\rm F}(\btheta)} f_i\,c_z(\bj,\mu_i),
\end{equation}
where $f_i$ is the fraction of observed galaxies in this sector that
were targeted in field $i$. Hence by identification of terms
\begin{equation}
c_z(\bj,\mu_i)= 0.99\, \left(1-\exp[\bj-\mu(\btheta)] \right) ,
\label{eqn:compl}
\end{equation}
where
\begin{equation}
\mu(\btheta)=-\ln\left[\sum_{i=1}^{N_{\rm F}(\btheta)}f_i\,\exp(-\mu_i)\right] .
\end{equation}

With this one can define the function
\begin{equation}
S(\btheta,\bj) = 
R(\btheta) \, c_z(\bj,\mu[\btheta]) / \bar c_z(\mu[\btheta])
\end{equation}
which is an estimate of the fraction of galaxies of apparent 
magnitude $\bj$ in the sector at position $\btheta$ that have redshift 
measurements. Here $\bar c_z(\mu[\btheta])$ is 
$c_z(\bj,\mu[\btheta])$ averaged over the expected 
apparent magnitude distribution of the galaxies in the sector.

Maps of $\bj^{\rm lim}(\btheta)$, $R(\btheta)$ and $\mu(\btheta)$
together with associated software are available for the 2dFGRS data in
the 100k Release (http://www.mso.anu.edu.au/2dFGRS/). Here, we
employ the method described in section~8 of Colless \etal
(\shortcite{colless01}) to generate these quantities for the more
extensive dataset used in this paper. The map of $R(\btheta)$ is shown
in Fig.~\ref{fig:mask} and the corresponding map of $\mu(\btheta)$ is
shown in Fig.~\ref{fig:mumask}.

\end{document}